\documentclass[letterpaper,twocolumn,twoside]{IEEEtran}
\usepackage{amsmath, amssymb, bm, cite, epsfig, psfrag}
\usepackage{algorithm,algorithmic}
\usepackage{ifthen}
\usepackage{subcaption}

\newfloat{algorithm}{tb}{lop}

\newboolean{anonymous}
\setboolean{anonymous}{false}
\newboolean{conference}
\setboolean{conference}{false}

\ifthenelse{\boolean{conference}}{
\addtolength\textfloatsep{-.3cm}
}{}%

\def\beq{\begin{equation}}
\def\eeq{\end{equation}}
\def\beqa{\begin{eqnarray}}
\def\eeqa{\end{eqnarray}}
\def\beqan{\begin{eqnarray*}}
\def\eeqan{\end{eqnarray*}}

\def\R{{\mathbb{R}}}

\def\argmax{\mathop{\mathrm{arg\,max}}}

\def\x{\times}

\newcommand{\leqn}[1]{{\overset{\mathrm{(#1)}}{\leq}}}

\newtheorem{definition}{Definition}
\newtheorem{theorem}{Theorem}

\newtheorem{assumption}{Assumption}

\def\xhat{\widehat{x}}

\def\zhat{\widehat{z}}

\def\taubar{\overline{\tau}}
\def\xihat{\widehat{\xi}}
\def\alphahat{\widehat{\alpha}}
\def\lambdabar{\overline{\lambda}}
\def\lambdahat{\widehat{\lambda}}
\def\thetabar{\overline{\theta}}

\def\PL{\mathrm{PL}}

\def\arr{\rightarrow}
\def\Exp{\mathbb{E}}
\def\var{\mbox{var}}
\def\cov{\mbox{cov}}
\def\tm1{t\! - \! 1}
\def\tp1{t\! + \! 1}

\newcommand{\pbf}{\mathbf{p}}

\newcommand{\rbf}{\mathbf{r}}

\newcommand{\sbf}{\mathbf{s}}

\newcommand{\vbf}{\mathbf{v}}

\newcommand{\wbf}{\mathbf{w}}

\newcommand{\xbf}{\mathbf{x}}
\newcommand{\xbfhat}{\widehat{\mathbf{x}}}

\newcommand{\ybf}{\mathbf{y}}
\newcommand{\zbf}{\mathbf{z}}

\newcommand{\zbfhat}{\widehat{\mathbf{z}}}
\newcommand{\Abf}{\mathbf{A}}

\newcommand{\Kbf}{\mathbf{K}}

\newcommand{\Phat}{\widehat{P}}

\newcommand{\Vbf}{\mathbf{V}}

\newcommand{\Xhat}{\widehat{X}}

\newcommand{\Zhat}{\widehat{Z}}

\def\xorc{\tilde{\mathbf{x}}}

\def\zorc{\tilde{\mathbf{z}}}
\def\porc{\tilde{\mathbf{p}}}
\def\sorc{\tilde{\mathbf{s}}}
\def\rorc{\tilde{\mathbf{r}}}
\def\tauorc{\tilde{\tau}}

\title{Approximate Message Passing with Consistent Parameter Estimation and Applications to Sparse Learning}

  \author{Ulugbek S. Kamilov, \IEEEmembership{Student Member,~IEEE},
          Sundeep Rangan, \IEEEmembership{Member,~IEEE},
          Alyson K. Fletcher, \IEEEmembership{Member,~IEEE},
          Michael Unser, \IEEEmembership{Fellow,~IEEE},
\thanks{The work of S. Rangan was supported by the National Science
Foundation under Grant No. 1116589. U. S. Kamilov and M. Unser were supported by the European Commission under Grant ERC-2010-AdG 267439-FUN-SP.}%
 \thanks{U.~S. Kamilov (email: ulugbek.kamilov@epfl.ch) is with Biomedical Imaging Group,
          {\'E}cole Polytechnique F{\'e}d{\'e}rale de Lausanne, Switzerland}
 \thanks{S. Rangan (email: srangan@poly.edu) is with
          Polytechnic Institute of New York University, Brooklyn, NY.}
  \thanks{A.~K. Fletcher (email: afletcher@ucsc.edu) is with
          the Department of Electrical Engineering,
          University of California, Santa Cruz.}
  \thanks{M. Unser (email: michael.unser@epfl.ch) is with Biomedical Imaging Group,
          {\'E}cole Polytechnique F{\'e}d{\'e}rale de Lausanne, Switzerland}%
}

\begin{document}

\markboth{Approximate Message Passing with Consistent Parameter Estimation}
        {Kamilov, Rangan, Fletcher and Unser}
\maketitle

\begin{abstract}
We consider the estimation of an i.i.d.\ (possibly non-Gaussian)
vector $\xbf \in \R^n$
from measurements $\ybf \in \R^m$
obtained by a general cascade model consisting
of a known linear transform followed by a probabilistic componentwise (possibly
nonlinear) measurement channel. A novel method, called
adaptive generalized approximate message passing
(Adaptive GAMP), that enables joint learning of the statistics of the
prior and measurement channel along with estimation
of the unknown vector $\xbf$ is presented.
The proposed algorithm is a generalization of a recently-developed EM-GAMP
that uses expectation-maximization (EM) iterations where the posteriors
in the E-steps are
computed via approximate message passing.
The methodology can be applied
to a large class of learning problems including the
learning of sparse priors in compressed sensing or identification
of linear-nonlinear cascade models in dynamical systems and neural spiking processes.
We prove that for large i.i.d.\ Gaussian transform matrices the asymptotic componentwise behavior of the adaptive GAMP
algorithm is predicted by a simple set of scalar state evolution equations.
In addition, we show that when a certain maximum-likelihood estimation
can be performed in each step, the adaptive GAMP method can yield
asymptotically consistent parameter estimates, 
which implies that the algorithm
achieves a reconstruction quality equivalent to the oracle algorithm that knows the correct parameter values.  Remarkably, this result applies to essentially
arbitrary parametrizations of the unknown distributions, including ones
that are nonlinear and non-Gaussian.
The adaptive GAMP methodology thus provides a systematic,
general and computationally efficient method
applicable to a large range of complex linear-nonlinear
models with provable guarantees.
\end{abstract}


\section{Introduction} \label{sec:intro}

\begin{figure}[t]
	\begin{minipage}[b]{1.0\linewidth}
		\centering
		\includegraphics[width=8.5cm]{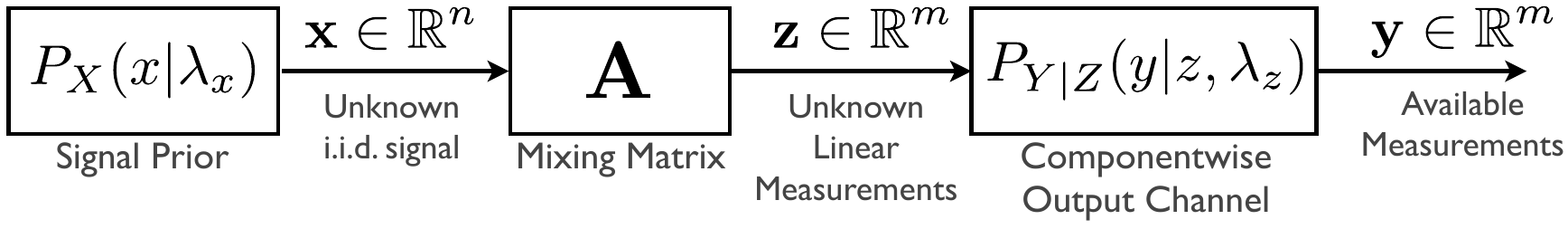}
	\end{minipage}
	\caption{Measurement model considered in this work.}
 	\label{fig:model}
\end{figure}

Consider the estimation of a random vector $\xbf \in \R^n$ from
the measurement model illustrated in Figure~\ref{fig:model}. The random vector $\xbf$, which is assumed to have independent and identically distributed (i.i.d.)\ components $x_j \sim P_X$,
is passed through a known linear transform that outputs $\zbf = \Abf\xbf$.
The components of $\ybf \in \R^m$ are generated by a componentwise transfer function $P_{Y|Z}$.
This work addresses the cases where the distributions $P_X$ and $P_{Y|Z}$
have some unknown parameters, $\lambda_x$ and $\lambda_z$, 
that must be learned in addition to the estimation of $\xbf$.

Such joint estimation and learning problems with linear transforms
and componentwise nonlinearities arise in a range of applications,
including empirical Bayesian approaches to
inverse problems in signal processing, linear regression and classification \cite{Tipping:01,West:03},
and, more recently, Bayesian compressed sensing
for estimation of sparse vectors $\xbf$
from underdetermined measurements\cite{WipfR:04,JiXueCarin:08,Cevher:09}.
Also, since the parameters in the output transfer function $P_{Y|Z}$ can model unknown nonlinearities, this problem formulation can be applied to the identification of linear-nonlinear cascade models of dynamical systems, in particular for neural spike responses \cite{BillingsFak:82,HunterKoren:86,SchwartzPRS2006}.

When the distributions $P_X$ and $P_{Y|Z}$ are known, or reasonably
bounded, there are a number of methods available that can be used for 
the estimation of the unknown vector $\xbf$.
In recent years, there has been significant interest in so-called
approximate message passing (AMP) and related methods 
based on Gaussian approximations of
loopy belief propagation (LBP) 
\cite{BoutrosC:02,TanakaO:05,GuoW:06,GuoW:07,
DonohoMM:09,DonohoMM:10-ITW1,DonohoMM:10-ITW2,
BayatiM:11,Rangan:10-CISS,Rangan:11-ISIT}.  These methods 
originate from CDMA multiuser detection problems in
\cite{BoutrosC:02,TanakaO:05,GuoW:06}, and have received considerable recent
attention in the context of compressed sensing \cite{DonohoMM:09,DonohoMM:10-ITW1,DonohoMM:10-ITW2,BayatiM:11,Rangan:10-CISS}.
See, also the survey article \cite{Montanari:12-bookChap}.
The Gaussian approximations used in AMP are also closely related to standard expectation propagation techniques \cite{Minka:01,Seeger:08},
but with additional simplifications that exploit the linear coupling between
the variables $\xbf$ and $\zbf$.
The key benefits of AMP methods are their computational simplicity, 
large domain of application,
and, for certain large random $\Abf$, their exact asymptotic performance characterizations
with testable conditions for optimality \cite{GuoW:06,GuoW:07,BayatiM:11,Rangan:10-CISS}.
This paper considers the so-called generalized AMP (GAMP) method of \cite{Rangan:11-ISIT}
that extends the algorithm in \cite{DonohoMM:09} to arbitrary output distributions $P_{Y|Z}$ (many original formulations assumed additive 
white Gaussian noise (AWGN) measurements).

However, although the current formulation of AMP and GAMP methods
is attractive conceptually, in practice, one often does not know the prior and noise distributions exactly.
To overcome this limitation, Vila and Schniter~\cite{VilaSch:11,VilaSch:12} and Krzakala \emph{et al.}~\cite{KrzMSSZ:11-arxiv,KrzMSSZ:12-arxiv} have 
recently proposed extension of AMP and GAMP based on Expectation 
Maximization (EM) that enable joint learning of the parameters $(\lambda_x,\lambda_z)$ along with the estimation of the vector $\xbf$.
While simulations indicate excellent performance, the analysis of these 
methods is difficult.
This work provides a unifying analytic framework for such AMP-based
joint estimation and learning methods. 
The main contributions of this paper are as follows:

\begin{itemize}

\item Generalization of the GAMP method of \cite{Rangan:11-ISIT} to 
    a class of algorithms we call \emph{adaptive GAMP} that
    enables joint estimation of the parameters $\lambda_x$ and $\lambda_z$ along with vector $\xbf$.
    The methods are computationally fast and general with potentially large domain of application.
    In addition, the adaptive GAMP methods include the EM-GAMP algorithms of
    \cite{VilaSch:11,VilaSch:12,KrzMSSZ:11-arxiv,KrzMSSZ:12-arxiv} 
    as special cases.

\item Exact characterization of the asymptotic behavior of adaptive GAMP. We show that, similar to the analysis of the AMP and GAMP algorithms in \cite{GuoW:06,GuoW:07,Rangan:10-CISS,Rangan:11-ISIT,BayatiM:11},
the componentwise asymptotic behavior of adaptive GAMP can be described exactly by a simple scalar \emph{state evolution} (SE) equations.

\item Demonstration of asymptotic consistency of adaptive GAMP with maximum-likelihood (ML) parameter estimation. Our main result shows that when the ML parameter estimation is computed exactly, the estimated parameters converge to the true values and the performance of adaptive GAMP asymptotically coincides with the performance of the oracle GAMP algorithm that knows the correct parameter values.
    Remarkably, this result applies to essentially arbitrary 
    parameterizations of the unknown distributions $P_X$ and $P_{Y|Z}$,
    thus enabling provably consistent estimation on non-convex and
    nonlinear problems.

\item Experimental evaluation of the algorithm for the problems of learning of sparse priors in compressed sensing and identification of linear-nonlinear cascade models in neural spiking processes. Our simulations illustrate the performance gain of adaptive GAMP and its asymptotic consistency. Adaptive GAMP thus provides a computationally-efficient method for a large class of joint estimation and learning problems
with a simple, exact performance characterization and provable conditions for asymptotic consistency.

\end{itemize}

\subsection{Related Literature}

As mentioned above,
the adaptive GAMP method proposed here can be seen as a generalization of the
EM methods in \cite{VilaSch:11,VilaSch:12,KrzMSSZ:11-arxiv,KrzMSSZ:12-arxiv}.
In \cite{VilaSch:11,VilaSch:12},
the prior $P_X$ is described by a generic $L$-term Gaussian mixture (GM) whose
parameters are identified by an EM procedure  \cite{DempLR:77}.  The
``expectation" or E-step is performed by
GAMP, which can approximately
determine the marginal posterior distributions of the components $x_j$ given the observations $\ybf$
and the current parameter estimates of the GM distribution $P_X$.  
A related EM-GAMP algorithm has also appeared in
\cite{KrzMSSZ:11-arxiv,KrzMSSZ:12-arxiv} for the case of certain sparse priors and AWGN outputs.
Simulations in \cite{VilaSch:11,VilaSch:12} show remarkably good performance
and computational speed for EM-GAMP
over a wide class of distributions, particularly in the context of compressed sensing. Also, using arguments from statistical physics, 
\cite{KrzMSSZ:11-arxiv,KrzMSSZ:12-arxiv} presents 
state evolution (SE) equations for the joint evolution
of the parameters and vector estimates  and confirms them numerically.   

As discussed in Section~\ref{sec:EM-GAMP},
EM-GAMP is a special case of adaptive GAMP with a particular choice of the
adaptation functions. Therefore,
one contribution of this paper is to provide a rigorous theoretical justification of the EM-GAMP methodology.
In particular, the current work provides a rigorous justification of the SE
 analysis in \cite{KrzMSSZ:11-arxiv,KrzMSSZ:12-arxiv}
 along with extensions to more general input and output channels and adaptation methods.  However, the methodology in \cite{KrzMSSZ:11-arxiv,KrzMSSZ:12-arxiv} in other ways is more general in that it can also study ``seeded" or ``spatially-coupled" matrices as proposed in
\cite{KrzMSSZ:11-arxiv,KrzMSSZ:12-arxiv,DonJavM:12}.  An interesting open question is whether the analysis methods in this paper can be extended to these scenarios as well.

An alternate method for joint learning and estimation has been presented 
in \cite{RanganFGS:12-ISIT}, which assumes that the distributions
on the source and output channels are themselves described by 
graphical models with the parameters $\lambda_x$ and $\lambda_z$
appearing as unknown variables.  
The method in \cite{RanganFGS:12-ISIT},
called Hybrid-GAMP, iteratively combines standard loopy BP with AMP methods.
One avenue of future work is to
see if methodology in this paper can be applied to analyze the Hybrid-GAMP
methods as well.

Finally, it should be pointed out that 
while simultaneous recovery of unknown parameters is appealing conceptually, it is not a strict requirement. An alternate solution to the problem is to assume that the signal belongs to a known class of distributions and to minimize the maximal mean-squared error (MSE) for the class. This minimax approach~\cite{DonohoJ:1994a} was proposed for AMP recovery of sparse signals in~\cite{DonohoMM:09}. Although minimax approach results in the estimators that are uniformly good over the entire class of distributions, there may be a significant gap between the MSE achieved by the minimax approach and the oracle algorithm that knows the distribution exactly. Indeed, this gap was
the main justification of the EM-GAMP methods in \cite{VilaSch:11,VilaSch:12}.
Due to its asymptotic consistency,
adaptive GAMP provably achieves the performance of the oracle algorithm.

\subsection{Outline of the Paper}

The paper is organized as follows: In Section~\ref{sec:gampReview}, we review the non-adaptive GAMP and corresponding state evolution equations. In Section~\ref{sec:gamp}, we present adaptive GAMP and describe ML parameter learning. In Section~\ref{sec:convProp}, we provide the main theorems characterizing asymptotic performance of adaptive GAMP and demonstrating its consistency. In Section~\ref{sec:numEx}, we provide numerical experiments demonstrating the applications of the method. Section~\ref{sec:concl} concludes the paper.
A conference version of this paper has appeared in \cite{KamRanFU:12-nips}.
This paper contains all the proofs, more detailed descriptions and additional
simulations.

\section{Review of GAMP} \label{sec:gampReview}

\subsection{GAMP Algorithm}

Before describing the adaptive GAMP algorithm, it is useful to review
the basic (non-adaptive) GAMP algorithm of \cite{Rangan:11-ISIT}.
Consider the estimation problem in Fig.~\ref{fig:model}
where the componentwise distributions on the inputs and outputs have
some parametric form,
\beq \label{eq:pxzlam}
    P_X(x|\lambda_x), \quad P_{Y|Z}(y|z, \lambda_z),
\eeq
where $\lambda_x \in \Lambda_x$ and $\lambda_z \in \Lambda_z$ represent
parameters of the distributions and $\Lambda_x$ and $\Lambda_z$
some parameter sets.

The GAMP algorithm of \cite{Rangan:11-ISIT} can be seen as a class of methods
for estimating the vectors $\xbf$ and $\zbf$ for the
case when the parameters $\lambda_x$ and $\lambda_z$ are \emph{known}.
In contrast, the adaptive GAMP method that is discussed in Section~\ref{sec:gamp} enables
\emph{joint} estimation of the parameters $\lambda_x$ and $\lambda_z$ along with the
vectors $\xbf$ and $\zbf$.
In order that to understand how the adaptation works,
it is best to describe the basic GAMP algorithm
as a special case of the more general adaptive GAMP procedure.

The basic GAMP algorithm corresponds to the special case of Algorithm~\ref{algo:gamp}
when the \emph{adaptation functions} $H_x^t$ and $H_z^t$ output fixed values
\beq \label{eq:Hfix}
    H^t_z(\pbf^t,\ybf,\tau^t_p) = \lambdabar^t_z, \qquad
    H^t_x(\rbf^t,\tau^t_r) = \lambdabar^t_x,
\eeq
for some \emph{pre-computed} sequence of parameters $\lambdabar^t_x$ and
$\lambdabar^t_z$.  By ``pre-computed", we mean that the values do not
depend on the data through the vectors $\pbf^t$, $\ybf^t$, and $\rbf^t$.
In the oracle scenario $\lambdabar^t_x$ and $\lambdabar^t_z$ are set to the true values
of the parameters and do not change with the iteration number $t$.


The \emph{estimation functions}
$G_x^t$, $G_z^t$ and $G_s^t$ determine the estimates for the
vectors $\xbf$ and $\zbf$, given the parameter values
$\lambdahat^t_x$ and $\lambdahat^t_z$.
As described in \cite{Rangan:11-ISIT}, there are two important sets of
choices for the estimation functions, resulting in two variants of GAMP:
\begin{itemize}
\item \emph{Sum-product GAMP:}  In this case, the estimation functions
are selected so that GAMP provides a Gaussian approximation of sum-product loopy
BP.  The estimates $\xbfhat^t$ and $\zbfhat^t$ then represent
approximations of the MMSE estimates of the vectors $\xbf$ and $\zbf$.

\item \emph{Max-sum GAMP:}  In this case, the estimation functions
are selected so that GAMP provides a quadratic approximation of max-sum loopy
BP and $\xbfhat^t$ and $\zbfhat^t$ represent
approximations of the MAP estimates.

\end{itemize}

The estimation functions
of the sum-product GAMP  are equivalent to
scalar MMSE estimation problems for the components of the vectors $\xbf$ and $\zbf$
observed in Gaussian noise. For max-sum GAMP, the estimation functions correspond
to scalar MAP problems.  Thus, for both versions, the GAMP method reduces
the vector-valued MMSE and MAP estimation problems to a sequence of scalar
AWGN problems combined with linear transforms by $\Abf$ and $\Abf^T$. GAMP
is thus computationally simple, with each iteration
involving only scalar nonlinear operations followed by linear transforms.
The operations are similar in form to separable and proximal minimization
methods widely used for such problems
\cite{WrightNF:09,ForGlow:83,GlowLeTal:89,HeLiaoHanY:02,WenGolYin:10}.
Appendix~\ref{sec:gampDetails} reviews the equations for the sum-product
GAMP.  More details, as well as the equations for max-sum GAMP 
can be found in \cite{Rangan:11-ISIT}.

\subsection{State Evolution Analysis} \label{sec:stateEvoGAMP}

In addition to its computational simplicity and generality,
a key motivation of the GAMP algorithm is that its asymptotic
behavior can be precisely characterized when $\Abf$ is a
large i.i.d.\ Gaussian transform.
The asymptotic behavior is described by
what is known as a \emph{state evolution} (SE) analysis.
By now, there are a large number of SE results for AMP-related
algorithms
\cite{BoutrosC:02,GuoW:06,GuoW:07,DonohoMM:09,DonohoMM:10-ITW1,
DonohoMM:10-ITW2,BayatiM:11,Rangan:10-CISS,Rangan:11-ISIT}.
Here, we review the particular SE analysis from \cite{Rangan:11-ISIT} 
which is based on the framework in \cite{BayatiM:11}.

\medskip
\begin{assumption}  \label{as:gamp}
Consider a sequence of random realizations of the GAMP algorithm,
indexed by the dimension $n$, satisfying the following assumptions:
\begin{itemize}
\item[(a)] \label{as:Gauss}
For each $n$, the matrix $\Abf \in \R^{m \x n}$ has
i.i.d.\ components with $A_{ij} \sim {\mathcal N}(0,1/m)$ and the dimension
$m=m(n)$ is a deterministic function of $n$ satisfying
$n/m \rightarrow \beta$ for some $\beta > 0$ as $n \rightarrow \infty$.

\item[(b)] \label{as:px0}
The input vectors $\xbf$
and initial condition $\xbfhat^0$
are deterministic sequences
whose components converge empirically
with bounded moments of order $s=2k-2$ as
\beq \label{eq:thetaxInit}
    \lim_{n \arr \infty} (\xbf, \xbfhat^0)
     \stackrel{\PL(s)}{=} (X,\Xhat^0),
\eeq
to some random vector $(X, \Xhat^0)$ for some $k \geq 2$.
\ifthenelse{\boolean{conference}}{See \cite[Appendix A]{KamRanF:12-supplement}}{
See Appendix \ref{sec:conv} }
for the precise definition of this form of convergence.

\item[(c)] \label{as:pyz}
The output vectors $\zbf$ and $\ybf \in \R^m$
are generated by
\beq \label{eq:zAx}
    \zbf = \Abf\xbf, \mbox{ and } y_i = h(z_i,w_i) \mbox{ for all } i=1,\ldots,m,
\eeq
for some scalar function $h(z,w)$ and vector $\wbf \in \R^m$
representing an output disturbance.
It is assumed that the
output disturbance vector $\wbf$ is deterministic,
but empirically converges as
\beq \label{eq:Wlim}
    \lim_{n \arr \infty} \wbf \stackrel{\PL(s)}{=} W,
\eeq
where $s=2k-2$ is as in Assumption~\ref{as:gamp} (b) and $W$ is some random
variable.  We let $P_{Y|Z}$ denote the conditional distribution of the random variable
$Y = h(Z,W)$.

\item[(d)] The estimation function $G_x^t(r,\tau_r,\lambda_x)$ and its derivative
with respect to $r$, is Lipschitz continuous in $r$ at
$(\tau_r,\lambda_x)=(\taubar_r^t,\lambdabar_x^t)$, where $\taubar_r^t$ is a deterministic parameter from the SE equations below. A similar assumptions holds for
$G_z^t(p,\tau_p,\lambda_z)$.

\item[(e)]
The adaptation functions $H_x^t$ and $H_z^t$
are set to \eqref{eq:Hfix} for
some deterministic sequence of parameters $\lambdabar_x^t$ and $\lambdabar_z^t$.
Also, in the estimation steps in lines \ref{line:zhat}, \ref{line:shat} and
\ref{line:xhat} of Algorithm~\ref{algo:gamp},
the values of the $\tau_p^t$ and $\tau_r^t$ are replaced with
the deterministic parameters
$\taubar_p^t$ and $\taubar_r^t$ from the SE equations defined below.
\end{itemize}
\end{assumption}

Assumption~\ref{algo:gamp}(a) simply states that we are considering
large, Gaussian i.i.d.\ matrices $\Abf$.
Assumptions (b) and (c) state that
the input vector $\xbf$ and output disturbance $\wbf$ are
modeled as deterministic, but whose empirical distributions asymptotically appear as i.i.d.
This deterministic model is one of key features of Bayati and Montanari's analysis
in \cite{BayatiM:11}.
Assumption (d) is a mild continuity condition.
Assumption (e) defines the restriction of adaptive GAMP to the non-adaptive
GAMP algorithm.  We will remove this final assumption later.

Note that, for now,
there is no assumption that the ``true" distribution of $X$ or
the true conditional distribution of $Y$ given $Z$ must belong to the
class of distributions \eqref{eq:pxzlam} for any parameters $\lambda_x$ and $\lambda_z$.
The analysis can thus model the effects of model mismatch.

Now, given the above assumptions, define the sets of vectors
\begin{subequations} \label{eq:thetaxz}
\beqa
    \theta_x^t &:=& \{(x_{j},r_j^t,\xhat_j^{\tp1}), j=1,\ldots,n \}, \\
    \theta_z^t &:=& \{(z_{i},\zhat_i^t,y_i,p_i^t), i=1,\ldots,m \}.
\eeqa
\end{subequations}
The first vector set, $\theta_x^t$, represents the
components of the the ``true," but unknown,
input vector $\xbf$, its GAMP estimate $\xbfhat^t$ as well as $\rbf^t$.
The second vector, $\theta_z^t$,
contains the components the ``true," but unknown,
output vector $\zbf$, its GAMP estimate $\zbfhat^t$, as well as $\pbf^t$
and the observed output $\ybf$.
The sets $\theta_x^t$ and
$\theta_z^t$ are implicitly functions of the dimension $n$.

The main result of \cite{Rangan:11-ISIT} shows that if we fix the
iteration $t$, and let $n \arr \infty$, the
asymptotic joint empirical distribution of the components of
these two sets $\theta_x^t$ and $\theta_z^t$ converges
to random vectors of the form
\beq \label{eq:thetabarxz}
    \thetabar_x^t := (X,R^t,\Xhat^{\tp1}), \qquad
       \thetabar_z^t := (Z,\Zhat^t,Y,P^t).
\eeq
We precisely state the nature of convergence momentarily (see Theorem~\ref{thm:stateEvoGAMP}).
In \eqref{eq:thetabarxz},
$X$ is the random variable in Assumption~\ref{as:gamp}(b), while
$R^t$ and $\Xhat^{\tp1}$ are given by
\begin{subequations} \label{eq:RVXt}
\begin{align}
    &R^t = \alpha^tX + V^t, \qquad
    V^t \sim {\mathcal N}(0,\xi_r^t), \\
   &\Xhat^{\tp1} = G_x^t(R^t,\taubar_r^t,\lambdabar_x^t)
\end{align}
\end{subequations}
for some deterministic constants $\alpha_r^t$, $\xi_r^t$, and $\taubar_r^t$ that are
defined below.  Similarly,
$(Z,P^t) \sim {\mathcal N}(0,\Kbf_p^t)$ for some covariance matrix $\Kbf_p^t$,
and
\beq \label{eq:YZPt}
    Y = h(Z,W), \qquad
    \Zhat^t = G_z^t(P^t,Y,\taubar_p^t,\lambdabar_z^t),
\eeq
where $W$ is the random variable in \eqref{eq:Wlim} and
$\Kbf_p^t$ contains deterministic constants.

The deterministic constants $\alpha_r^t$, $\xi_r^t$, $\taubar_r^t$ and $\Kbf_p^t$
represent parameters of the distributions of $\thetabar_x^t$ and $\thetabar_z^t$
and depend on both the distributions of the input and outputs
as well as the choice of the estimation and adaptation functions.
The SE equations provide a simple method for recursively computing
these parameters.  The equations are best described
algorithmically as shown in Algorithm~\ref{algo:SE}.
In order that we do not repeat ourselves, in
Algorithm~\ref{algo:SE}, we have written the SE equations for adaptive GAMP:
For non-adaptive GAMP,
the updates \eqref{eq:lambarzSE} and \eqref{eq:lambarxSE} can be ignored as the
values of $\lambdabar_z^t$ and $\lambdabar_x^t$ are pre-computed.

With these definitions, we can state the main result from \cite{Rangan:11-ISIT}.

\medskip
\begin{theorem}[\cite{Rangan:11-ISIT}] \label{thm:stateEvoGAMP}
Consider the random vectors $\theta_x^t$ and
$\theta_z^t$ generated by the outputs of GAMP under
Assumption \ref{as:gamp}.  Let $\thetabar_x^t$ and $\thetabar_z^t$
be the random vectors in \eqref{eq:thetabarxz} with the parameters
determined by the SE equations in Algorithm~\ref{algo:SE}.
Then, for any fixed $t$,
the components of $\theta_x^t$ and
$\theta_z^t$ converge empirically with bounded moments of order $k$ as
\beq \label{eq:thetaLimGAMP}
    \lim_{n \arr \infty} \theta_x^t \stackrel{\PL(k)}{=}
        \thetabar_x^t, \qquad
    \lim_{n \arr \infty} \theta_z^t \stackrel{\PL(k)}{=}
        \thetabar_z^t.
\eeq
where $\thetabar_x^t$ and $\thetabar_z^t$ are given in \eqref{eq:thetabarxz}.
In addition, for any $t$, the limits
\beq \label{eq:tauLimGAMP}
 \lim_{n \arr \infty} \tau_r^t = \taubar_r^t, \qquad
 \lim_{n \arr \infty} \tau_p^t = \taubar_p^t,
\eeq
also hold almost surely.
\end{theorem}

The theorem shows that the behavior of any component of the vectors
$\xbf$ and $\zbf$ and their GAMP estimates $\xbfhat^t$ and $\zbfhat^t$ are
distributed identically to a simple scalar equivalent system with random variables
$X$, $Z$, $\Xhat^t$ and $\Zhat^t$.
This scalar equivalent model appears in several analyses
and can be thought of as a \emph{single-letter characterization} \cite{BaronSB:10} of the system.
Remarkably, this limiting property holds for essentially arbitrary distributions
and estimation functions, even ones that arise from problems that are highly nonlinear
or noncovex.
From the single-letter characterization, one can compute the
asymptotic value of essentially any componentwise performance metric such
as mean-squared error or detection accuracy.  
Similar single-letter characterizations can also be derived by 
arguments from statistical physics 
\cite{Tanaka:02,GuoV:05,CaireSTV:11,KrzMSSZ:11-arxiv,RanganFG:12-IT}.

\section{Adaptive GAMP} \label{sec:gamp}

\begin{algorithm}
\caption{Adaptive GAMP}
\begin{algorithmic}[1]  \label{algo:gamp}
\REQUIRE{ Matrix $\Abf$, estimation functions $G_x^t$,
$G_s^t$ and $G_z^t$ and adaptation functions $H_x^t$
and $H_z^t$. }

\STATE{ Set $t \gets 0$, $\sbf^{t-1} \gets 0$ and select some
initial values for $\xbfhat^0$ and  $\tau_x^0$.  }
\REPEAT

    \STATE{ } \COMMENT{Output node update}
    \STATE{ $\tau_p^t \gets \|\Abf\|^2_F\tau_x^t/m$ }
        \label{line:taup}
    \STATE{ $\pbf^t \gets \Abf\xbfhat^t - \sbf^{\tm1}\tau_p^t$ }
        \label{line:phat}
    \STATE{ $\lambdahat_z^t \gets H_z^t(\pbf^t,\ybf,\tau_p^t)$ }
        \label{line:lamz}
    \STATE{ $\zhat_i^t \gets G_z^t(p_i^t,y_i,\tau_p^t,\lambdahat_z^t)$ for all $i=1,\ldots,m$ }
        \label{line:zhat}
    \STATE{ $s_i^t \gets G_s^t(p_i^t,y_i,\tau_p^t,\lambdahat_z^t)$ for all $i=1,\ldots,m$ }
        \label{line:shat}
    \STATE{ $\tau_s^t \gets -(1/m)\sum_i  \partial G_s^t(p_i^t,y_i,\tau_p^t,\lambdahat_z^t) / \partial p_i^t$ }
        \label{line:taus}

    \STATE{ }
    \STATE{ } \COMMENT{Input node update}
   \STATE{ $1/\tau_r^t \gets \|\Abf\|^2_F\tau_s^t/n$ }
         \label{line:taur}
    \STATE{ $\rbf^t = \xbf^t + \tau_r^t\Abf^\mathrm{T}\sbf^t$ }
        \label{line:rhat}
    \STATE{ $\lambdahat_x^t \gets H_x^t(\rbf^t,\tau_r^t)$ }
        \label{line:lamx}
    \STATE{ $\xhat^{\tp1}_j \gets G_x^t(r_j^t,\tau_r^t,\lambdahat_x^t)$
        for all $j=1,\ldots,n$ }    \label{line:xhat}
    \STATE{ $\tau_x^{\tp1} \gets
        (\tau_r^t/n)\sum_j  \partial G_x^t(r_j^t,\tau_r^t,\lambdahat_x^t) /\partial r_j$ } \label{line:taux}

\UNTIL{Terminated}

\end{algorithmic}
\end{algorithm}

As described in the previous section,
the standard GAMP algorithm of \cite{Rangan:11-ISIT} 
considers the case when the parameters $\lambda_x$
and $\lambda_z$ in the distributions in \eqref{eq:pxzlam} are
known.
The adaptive GAMP method proposed in this paper,
and shown in Algorithm~\ref{algo:gamp}, is an extension of the standard
GAMP procedure that enables simultaneous identification
of the parameters  $\lambda_x$ and $\lambda_z$
along with estimation of the vectors $\xbf$ and $\zbf$.
The key modification is
the introduction of the two \emph{adaptation functions}:
$H_z^t(\pbf^t,\ybf,\tau_p^t)$ and $H_x^t(\rbf^t,\tau_r^t)$.
In each iteration, these functions output estimates, $\lambdahat_z^t$
and $\lambdahat_x^t$ of the parameters based on the data
$\pbf^t$, $\ybf$, $\rbf^t$, $\tau_p^t$ and $\tau_r^t$.
We saw the standard GAMP method
corresponds to the adaptation functions in \eqref{eq:Hfix}
which outputs fixed values $\lambdabar_z^t$
and $\lambdabar_x^t$ that do not depend on the data, and can be used when
the true parameters are known.  For the case when the true parameters
are not known, we will see that a simple maximum likelihood (ML)
can be used to estimate the parameters from the data.

\subsection{ML Parameter Estimation} \label{sec:MLadapt}


To understand how to estimate parameters via the adaptation functions,
observe that from Theorem~\ref{thm:stateEvoGAMP}, we know that
the distribution of the components of $\rbf^t$ are distributed identically to the
scalar $R^t$ in \eqref{eq:RVXt}.  Now, the distribution of $R^t$
only depends on three parameters --
$\alpha_r^t$, $\xi_r^t$ and $\lambda_x$.  It is thus natural to attempt to estimate
these parameters from the empirical distribution of the components of $\rbf^t$.

To this end, let $\phi_x(r,\lambda_x,\alpha_r,\xi_r)$ be the log likelihood
\beq \label{eq:phixLL}
    \phi_x(r,\lambda_x,\alpha_r,\xi_r) = \log p_R(r|\lambda_x,\alpha_r,\xi_r),
\eeq
where the right-hand side is the probability density of a random variable
$R$ with distribution
\begin{equation*}
    R = \alpha_rX + V, \quad X\sim P_X(\cdot|\lambda_x), \quad
    V \sim {\mathcal N}(0,\xi_r).
\end{equation*}
Then, at any iteration $t$, we can
attempt to perform a maximum-likelihood (ML) estimate
\beqa
    \lefteqn{ \lambdahat^t_x =  H_x^t(\rbf^t,\tau_r^t) }\nonumber \\
    & & = \argmax_{\lambda_x \in \Lambda_x}  \max_{(\alpha_r,\xi_r) \in S_x(\tau_r^t)}
    \left\{
        \frac{1}{n} \sum_{j=1}^{n} \phi_x(r^t_j,\lambda_x,\alpha_r,\xi_r)\right\}.
        \hspace{0.5cm}         \label{eq:HxML}
\eeqa
Here, the set $S_x(\tau_r^t)$ is a set of possible values for the parameters
$\alpha_r,\xi_r$.  The set may depend on the measured variance $\tau_r^t$.
We will see the precise role of this set below.

Similarly, the joint distribution of the components of $\pbf^t$ and $\ybf$
are distributed according to the scalar $(P^t,Y)$ which depend only
on the parameters $\Kbf_p$ and $\lambda_z$.  Thus, we can define the likelihood
\begin{equation} \label{eq:phizLL}
    \phi_z(p,y,\lambda_z,\Kbf_p) = \log p_{P,Y} (p,y|\lambda_z,\Kbf_p),
\end{equation}
where the right-hand side is the joint
probability density of $(P,Y)$ with distribution
\begin{equation*}
    Y \sim P_{Y|Z}(\cdot|Z,\lambda_z), \quad (Z,P) \sim {\mathcal N}(0,\Kbf_p).
\end{equation*}
Then, we can attempt to estimate $\lambda_z$ via the ML estimate
\begin{align}
  \lambdahat^t_z &=  H_z^t(\pbf^t,\ybf,\tau_p^t) \nonumber\\
    &= \argmax_{\lambda_z \in \Lambda_z}  \max_{\Kbf_p \in S_z(\tau_p^t)} \left\{
        \frac{1}{m} \sum_{i=1}^{m} \phi_z(p^t_i,y_i,\Kbf_p)\right\}.
        \label{eq:HzML}
\end{align}
Again, the set $S_z(\tau_p^t)$ is a set of possible covariance matrices
$\Kbf_p$.

\subsection{Relation to EM-GAMP}  \label{sec:EM-GAMP}

It is useful to briefly compare the above ML parameter estimation
with the EM-GAMP method proposed by Vila and Schniter in
 \cite{VilaSch:11,VilaSch:12} and Krzakala \emph{et.\ al.\ }  in \cite{KrzMSSZ:11-arxiv,KrzMSSZ:12-arxiv}.
Both of these methods combine the Bayesian AMP
\cite{DonohoMM:10-ITW1,DonohoMM:10-ITW2} or GAMP algorithms \cite{Rangan:11-ISIT}
with a standard EM procedure \cite{DempLR:77} as follows.
First, the algorithms use the sum-product version of the AMP / GAMP algorithms,
so that the outputs can provide an estimate of the posterior distributions on the
components of $\xbf$ given the current parameter estimates.
Specifically, at any iteration $t$, define the distribution
\beqa
    \lefteqn{ \Phat^t_j(x_j|r^t_j,\tau_r^t,\lambdahat_x^{\tm1}) } \nonumber \\
    &=& \frac{1}{Z}
    \exp\left[ -\frac{1}{2\tau_r^t}|x_j-r^t_j|^2\right]P_X(x_j|\lambdahat_x^{\tm1}).
    \label{eq:PhatPost}
\eeqa
For the sum-product AMP or GAMP algorithms, it is shown in \cite{Rangan:11-ISIT}
that the SE equations simplify so that $\alpha_r^t=1$ and $\xi_r^t = \taubar_r^t$,
if the parameters were selected correctly.
Therefore, from Theorem~\ref{thm:stateEvoGAMP}, the conditional distribution
$P(x_j|r_j^t)$ should approximately match the distribution \eqref{eq:PhatPost}
for large $n$.
If, in addition, we treat $r_j^t$ and $\tau_r^t$ as sufficient statistics for estimating
$x_j$ given $\ybf$ and $\Abf$, then $\Phat^t_j$
can be treated as an approximation for
the posterior distribution of $x_j$ given the current parameter estimate $\lambdahat_x^{\tm1}$.
Some justification for this last step can be found in \cite{GuoW:06,GuoW:07,Rangan:10-CISS}.
Using the approximation, we can approximately implement the EM procedure
to update the parameter estimate via a maximization
\beqa
    \lefteqn{ \lambdahat^t_x = H^t_x(\rbf^t,\tau^t_r)  }
        \nonumber \\
    &:=& \argmax_{\lambda_x \in \Lambda_x}
     \frac{1}{n} \sum_{j=1}^n \Exp \Bigl[ \log P_X(x_j|\lambda_x) | \Phat^t_j \Bigr],
     \label{eq:lamEM}
\eeqa
where the expectation is with respect to the distribution in \eqref{eq:PhatPost}.
In \cite{VilaSch:11,VilaSch:12}, the parameter update \eqref{eq:lamEM}
is performed only once every few iterations to allow $\Phat^t$
to converge to the approximation of the posterior distribution of $x_j$
given the current parameter estimates.  In \cite{KrzMSSZ:11-arxiv,KrzMSSZ:12-arxiv},
the parameter estimate is updated every iteration.
A similar procedure can be performed for the estimation of $\lambda_z$.

We thus see that the EM-GAMP procedures in
\cite{VilaSch:11,VilaSch:12} and in
\cite{KrzMSSZ:11-arxiv,KrzMSSZ:12-arxiv} are both special cases of the
adaptive GAMP algorithm in Algorithm~\ref{algo:gamp} with particular
choices of the adaptation functions $H^t_x$ and  $H^t_z$.
As a result, our analysis in Theorem~\ref{thm:stateEvo} below
can be applied to these algorithms to provide rigorous asymptotic
characterizations of the EM-GAMP performance.
However, at the current time, we can only prove
the asymptotic consistency result, Theorem~\ref{thm:consistent},
for the ML adaptation functions \eqref{eq:HxML} and \eqref{eq:HzML}
described above.

That being said, it should be pointed out that EM-GAMP
update \eqref{eq:lamEM} is generally computationally much simpler than
the ML updates
in \eqref{eq:HxML} and \eqref{eq:HzML}.  For example, when $P_X(x|\lambda_x)$
is an exponential family, the optimization in \eqref{eq:lamEM} is convex.
Also, the optimizations in \eqref{eq:HxML} and \eqref{eq:HzML} require searches over
additional parameters such as $\alpha_r$ and $\xi_r$.
Thus, an interesting avenue of future work is to apply the analysis result,
Theorem~\ref{thm:consistent} below, to see if the EM-GAMP method or some similarly
computationally simple technique can be developed which also provides
asymptotic consistency.

\section{Convergence and Asymptotic Consistency with Gaussian Transforms} \label{sec:convProp}

\subsection{General State Evolution Analysis} \label{sec:stateEvo}

\begin{algorithm}
\caption{Adaptive GAMP State Evolution} \label{algo:SE}
Given the distributions in Assumption~\ref{as:gamp}, compute the sequence
of parameters as follows:
\begin{itemize}
\item \emph{Initialization} Set $t=0$ with
\beq \label{eq:SEinit}
    \Kbf_x^0 = \cov(X, \Xhat^0), \qquad \taubar_x^0 = \tau_x^0,
\eeq
where the expectation is over the random variables $(X,\Xhat^0)$ in Assumption \ref{as:gamp}(b)
and $\tau_x^0$ is the initial value in the GAMP algorithm.
\item \emph{Output node update:} Compute the variables associated with the
output nodes
\begin{subequations} \label{eq:outSE} Compute the variables
\beqa
    \taubar_p^t &=& \beta\taubar_x^t, \qquad \Kbf_p^t = \beta \Kbf_x^t, \\
    \lambdabar_z^t &=& H_z^t(P^t,Y, \taubar_p^t), \label{eq:lambarzSE} \\
    \taubar_r^t &=& -\Exp^{-1}\left[ \frac{\partial}{\partial p}
        G_s^t(P^t,Y,\taubar_p^t,\lambdabar_z^t) \right], \label{eq:taubarrSE}\\
    \xi_r^t &=& (\taubar_r^t)^2\Exp\left[ G_s^t(P^t,Y,\taubar_p^t,\lambdabar_z^t) \right], \\
    \alpha_r^t &=& \taubar_r^t \Exp\left[ \frac{\partial}{\partial z}
        G_s^t(P^t,h(Z,W),\taubar_p^t,\lambdabar_z^t) \right],
        \label{eq:alphaSE}
\eeqa
\end{subequations}
where the expectations are over the random variables $(Z,P^t) \sim {\mathcal N}(0,\Kbf_p^t)$
and $Y$ is given in \eqref{eq:YZPt}.
\item \emph{Input node update:} Compute
\begin{subequations} \label{eq:inSE}
\beqa
    \lambdabar_x^t &=& H_x^t(R^t,\taubar_r^t), \label{eq:lambarxSE} \\
    \taubar_x^{\tp1} &=& \taubar_r^t\Exp\left[ \frac{\partial}{\partial r}
        G_x^t(R^t,\taubar_r^t,\lambdabar_x^t) \right], \\
    \Kbf^{\tp1}_x &=& \cov(X,\Xhat^{\tp1}),
\eeqa
\end{subequations}
where the expectations are over the random variables in \eqref{eq:RVXt}.
\end{itemize}
\end{algorithm}

Before proving the asymptotic consistency of the adaptive GAMP method
with ML adaptation,
we first prove the following more general convergence result.

\begin{assumption}  \label{as:agamp}
Consider the adaptive GAMP algorithm
running on a sequence of problems indexed by the dimension $n$, satisfying
the following assumptions:
\begin{itemize}
\item[(a)]  Same as Assumption~\ref{as:gamp}(a) to (c) with $k=2$.

\item[(b)]  For every $t$,
\ifthenelse{\boolean{conference}}{
the adaptation functions $H_x^t(\rbf,\tau_r)$ and $H_z^t(\ybf,\pbf,\tau_p)$
are weakly pseudo-Lipschitz continuous in the vectors $\rbf$ and $(\ybf,\pbf)$
and continuous in the scalars $\tau_r$ and $\tau_p$ -- see
\cite{KamRanF:12-supplement} for details}{
the adaptation function $H_x^t(\rbf,\tau_r)$ can be regarded as a functional
over $\rbf$ satisfying the following weak pseudo-Lipschitz continuity property:
Consider any sequence of vectors $\rbf = \rbf^{(n)}$
and sequence of scalars $\tau_r = \tau_r^{(n)}$, indexed by $n$
satisfying
\[
    \lim_{n \arr \infty} \rbf^{(n)} \stackrel{\PL(k)}{=} R^t, \qquad
    \lim_{n \arr \infty} \tau_r^{(n)} = \taubar_r^t,
\]
where $R^t$ and $\taubar_r^t$ are the outputs of the state evolution equations
defined below.  Then,
\[
    \lim_{n \arr \infty} H_x^t(\rbf^{(n)},\tau_r^{(n)}) = H_x^t(R^t,\taubar_r^t).
\]
Similarly, $H_z^t(\ybf,\pbf,\tau_p)$ satisfies analogous continuity conditions
in $\tau_p$ and $(\ybf,\pbf)$.  See Appendix \ref{sec:conv} for a general
definition of weakly pseudo-Lipschitz continuous functionals. }

\item[(c)] \ifthenelse{\boolean{conference}}{
The estimation functions $G_x^t(r,\tau_r,\lambda_x)$,
$G_z^t(p,y,\tau_p,\lambda_z)$ and $G_z^t(p,y,\tau_p,\lambda_z)$
and their derivatives with respects to $r$ and $p$ are
continuous in $\lambda_x$ and $\lambda_z$ uniformly over the other parameters,
and Lipschitz continuous in $r$ and $p$ -- see \cite{KamRanF:12-supplement} for details.
}{The scalar function $G_x^t(r,\tau_r,\lambda_x)$
and its derivative ${G'}_x^t(r,\tau_r,\lambda_x)$ with respect to $r$
are continuous in $\lambda_x$
uniformly over $r$ in the following sense:
For every $\epsilon > 0$, $t$, $\tau_r^*$ and $\lambda_x^* \in \Lambda_x$,
there exists an open neighborhood $U$ of $(\tau_r^*,\lambda_x^*)$
such that for all $(\tau_r,\lambda_x) \in U$ and $r$,
\begin{align*}
    &|G_x^t(r,\tau_r,\lambda_x)-G_x^t(r,\tau^*_r,\lambda^*_x)| < \epsilon, \\
    &|{G'}_x^t(r,\tau_r,\lambda_x)-{G'}_x^t(r,\tau^*_r,\lambda^*_x)| < \epsilon.
\end{align*}
In addition, the functions $G_x^t(r,\tau_r,\lambda_x)$
and  ${G'}_x^t(r,\tau_r,\lambda_x)$ must be Lipschitz continuous in $r$
with a Lipschitz constant that can be selected continuously in $\tau_r$ and $\lambda_x$.
The functions $G_s^t(p,y,\tau_p,\lambda_z)$, $G_z^t(p,y,\tau_p,\lambda_z)$
and their derivatives ${G'}_s^t(p,y,\tau_p,\lambda_z)$
${G'}_z^t(p,y,\tau_p,\lambda_z)$ satisfy analogous
continuity assumptions with respect to $p$, $y$, $\tau_p$ and $\lambda_z$.
}
\end{itemize}
\end{assumption}

\ifthenelse{\boolean{conference}}{See \cite{KamRanF:12-supplement} for further discussion.}{
Assumptions (b) and (c) are somewhat technical, but mild, continuity conditions
that can be satisfied by a large class of adaptation functionals and estimation
functions.
For example, from the definitions in Appendix \ref{sec:conv}, the continuity assumption (d)
will be satisfied for any functional given by an empirical average
\[
    H^t_x(\rbf,\tau_r) = \frac{1}{n} \sum_{j=1}^n \phi^t_x(r_j,\tau_r),
\]
where, for each $t$, $\phi^t_x(r_j,\tau_r)$ is pseudo-Lipschitz continuous in $r$ of order $p$
and continuous in $\tau_r$ uniformly over $r$.  A similar functional can be used for $H^t_z$.
As we will see in Section \ref{sec:consistent}, the ML functionals
\eqref{eq:HxML} and \eqref{eq:HzML}
will also satisfy the conditions of this assumption.
}
\begin{theorem} \label{thm:stateEvo}
Consider the random vectors $\theta_x^t$ and
$\theta_z^t$ generated by the outputs of the adaptive GAMP under
Assumption \ref{as:agamp}.  Let $\thetabar_x^t$ and $\thetabar_z^t$
be the random vectors in \eqref{eq:thetabarxz} with the parameters
determined by the SE equations in Algorithm~\ref{algo:SE}.
Then, for any fixed $t$,
the components of $\theta_x^t$ and
$\theta_z^t$ converge empirically with bounded moments of order $k=2$ as
\beq \label{eq:thetaLim}
    \lim_{n \arr \infty} \theta_x^t \stackrel{\PL(k)}{=}
        \thetabar_x^t, \qquad
    \lim_{n \arr \infty} \theta_z^t \stackrel{\PL(k)}{=}
        \thetabar_z^t.
\eeq
where $\thetabar_x^t$ and $\thetabar_z^t$ are given in \eqref{eq:thetabarxz}.
In addition, for any $t$, the limits
\begin{subequations} \label{eq:taulamLim}
\begin{align} 
   & \lim_{n \arr \infty} \lambda_x^t = \lambdabar_x^t, \qquad
    &\lim_{n \arr \infty} \lambda_z^t = \lambdabar_z^t, \\
   & \lim_{n \arr \infty} \tau_r^t = \taubar_r^t, \qquad
   &\lim_{n \arr \infty} \tau_p^t = \taubar_p^t,
\end{align}
\end{subequations}
also hold almost surely.
\end{theorem}
\ifthenelse{\boolean{conference}}{}{
\emph{Proof:} See Appendix \ref{sec:stateEvoPf}.
}

The result is a natural generalization of Theorem~\ref{thm:stateEvoGAMP}
and provides a simple extension of the SE analysis to incorporate the adaptation.
The SE analysis applies to essentially arbitrary adaptation functions.
It particular, it can be used to analyze both the behavior of the
adaptive GAMP algorithm with either ML and EM-GAMP adaptation
functions in the previous section.

The proof is straightforward and is based on a continuity argument
also used in \cite{RanganF:12-ISIT}.

\subsection{Asymptotic Consistency with ML Adaptation} \label{sec:consistent}

We cam now use Theorem~\ref{thm:stateEvo} to prove
the asymptotic consistency of the adaptive GAMP method with the
ML parameter estimation described in Section~\ref{sec:MLadapt}.
The following
two assumptions can be regarded as \emph{identifiability} conditions.

\begin{definition} \label{def:Pxident}  Consider a family of distributions,
$\{P_X(x|\lambda_x), \lambda_x \in \Lambda_x\}$, a set $S_x$
of parameters $(\alpha_r,\xi_r)$ of a Gaussian channel and
function $\phi_x(r,\lambda_x,\alpha_r,\xi_r)$.
We say that $P_X(x|\lambda_x)$ is \emph{identifiable with Gaussian outputs} with
parameter set $S_x$ and function $\phi_x$ if:
\begin{itemize}
\item[(a)] The sets $S_x$ and $\Lambda_x$ are compact.
\item[(b)] For any ``true" parameters $\lambda_x^* \in \Lambda_x$,
and $(\alpha_r^*,\xi_r^*) \in S_x$, the maximization
\beqa
    \lefteqn{ \lambdahat_x = \argmax_{\lambda_x \in \Lambda_x}
    \max_{(\alpha_r,\xi_r) \in S_x}  }
        \nonumber \\
    & & \Exp\left[ \phi_x(\alpha_r^* X+V,\lambda_x,\alpha_r,\xi_r) |
        \lambda_x^*, \xi_r^* \right],
        \label{eq:lamxScore}
\eeqa
is well-defined, unique and returns the true value, $\lambdahat_x=\lambda_x^*$.
The expectation in \eqref{eq:lamxScore} is with respect to 
$X \sim P_X(\cdot|\lambda_x^*)$
and $V \sim {\mathcal N}(0,\xi_r^*)$.
\item[(c)]  For every $\lambda_x$ and $\alpha_r$, $\xi_r$, the
function $\phi_x(r,\lambda_x,\alpha_r,\xi_r)$ is pseudo-Lipschitz continuous of
order $k=2$ in $r$.
\ifthenelse{\boolean{conference}}{
In addition, it is continuous in $\lambda_x,\alpha_r,\xi_r$ uniformly over $r$ --
see \cite{KamRanF:12-supplement} for details.}{
In addition, it is continuous in $\lambda_x,\alpha_r,\xi_r$ uniformly over $r$ in the following
sense:  For every $\epsilon > 0$ and $\lambdahat_x,\alphahat_r,\xihat_r$,
there exists an open neighborhood $U$ of $\lambdahat_x,\alphahat_r,\xihat_r$,
such that for all $(\lambda_x,\alpha_r,\xi_r) \in U$ and all $r$,
\[
    | \phi_x(r,\lambda_x,\alpha_r,\xi_r) - \phi_x(r,\lambdahat_x,\alphahat_r,\xihat_r)|
        < \epsilon.
\]
}
\end{itemize}
\end{definition}

\begin{definition} \label{def:Pzident}  Consider a family of conditional distributions,
$\{P_{Y|Z}(y|z,\lambda_z), \lambda_z \in \Lambda_z\}$ generated by
the mapping $Y=h(Z,W,\lambda_z)$ where $W \sim P_W$ is some random variable
and $h(z,w,\lambda_z)$ is a scalar function.  Let $S_z$ be a set of
covariance matrices $\Kbf_p$ and let $\phi_z(y,p,\lambda_z,\Kbf_p)$
  be some function.  We say that conditional distribution family
$P_{Y|Z}(\cdot|\cdot,\lambda_z)$ is \emph{identifiable with Gaussian inputs}
with covariance set $S_z$ and function $\phi_z$ if:
\begin{itemize}
\item[(a)] The parameter sets $S_z$ and $\Lambda_z$ are compact.
\item[(b)] For any ``true" parameter $\lambda_z^* \in \Lambda_z$
and true covariance $\Kbf^*_p$, the maximization
\beqa
    \lefteqn{ \lambdahat_z = \argmax_{\lambda_z \in \Lambda_z} \max_{\Kbf_p \in S_z} }
    \nonumber \\
    & & \Exp\left[ \phi_z(Y,P,\lambda_z,\Kbf_p) | \lambda_z^*, \Kbf_p^* \right],  \label{eq:lamzScore}
\eeqa
is well-defined, unique and returns the true value, $\lambdahat_z=\lambda_z^*$,
The expectation in \eqref{eq:lamzScore} is with respect to
$Y|Z \sim P_{Y|Z}(y|z,\lambda_z^*)$ and $(Z,P) \sim {\mathcal N}(0,\Kbf_p^*)$.
\item[(c)]  For every $\lambda_z$ and $\Kbf_p$, the function
$\phi_z(y,p,\lambda_z,\Kbf_p)$
is pseudo-Lipschitz continuous in $(p,y)$ of order $k=2$.
In addition, it is continuous in $\lambda_p,\Kbf_p$ uniformly over $p$ and $y$.
\end{itemize}
\end{definition}

Definitions \ref{def:Pxident} and \ref{def:Pzident} essentially require that the
parameters $\lambda_x$ and $\lambda_z$ can be identified
through a maximization.  The functions $\phi_x$
and $\phi_z$ can be the log likelihood functions \eqref{eq:phixLL} and
\eqref{eq:phizLL}, although we permit other functions as well, since the maximization
may be computationally simpler.  Such functions are sometimes called
\emph{pseudo-likelihoods}.
The existence of a such a function is a mild condition.
Indeed, if such a function does not exists, then the distributions on $R$ or $(Y,P)$
must be the same for at least two different parameter values.
In that case, one cannot hope to identify the correct value from observations of
the vectors $\rbf^t$ or $(\ybf,\pbf^t)$.

\begin{figure*}[t]
\centering
	\includegraphics[width=17cm]{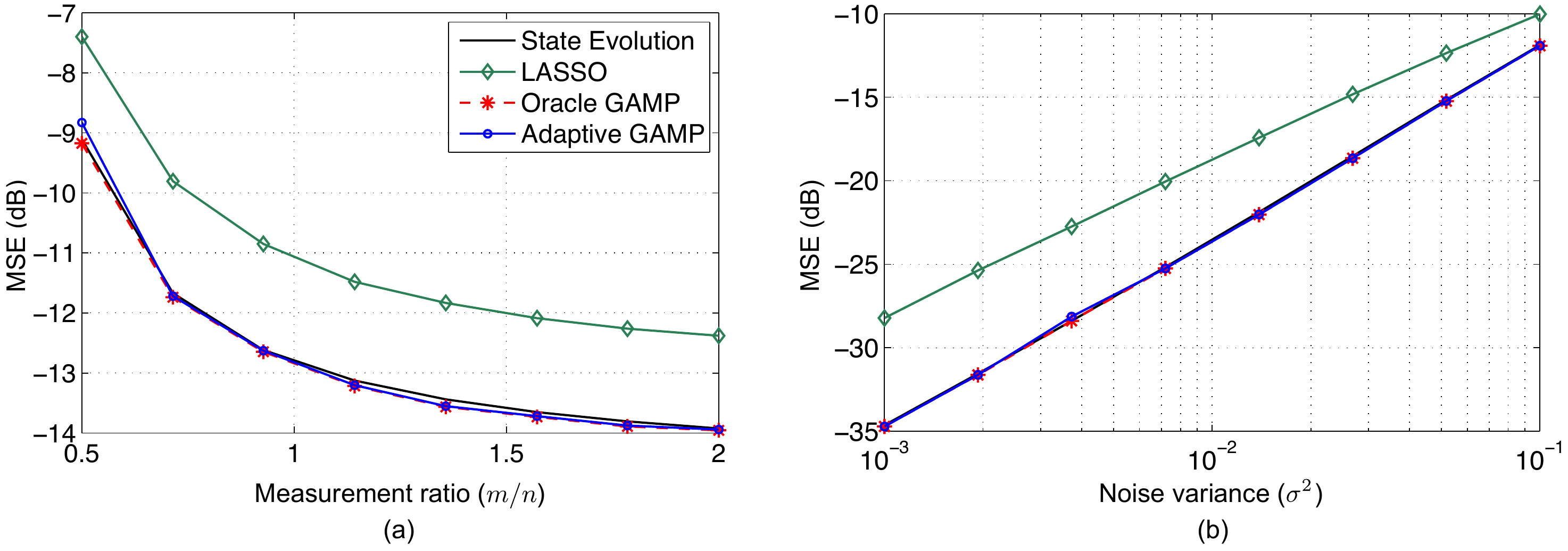}
	\caption{Reconstruction of a Gauss-Bernoulli signal from noisy measurements. The average reconstruction MSE is plotted against (a) measurement ratio $m/n$ and (b) AWGN variance $\sigma^2$. The plots illustrate that adaptive GAMP yields considerable improvement over $\ell_1$-based LASSO estimator. Moreover, it exactly matches the performance of oracle GAMP that knows the prior parameters.}
	\label{fig:mseresults}
\end{figure*}

\begin{assumption} \label{as:agamp-ML}
Let $P_X(x|\lambda_x)$ and $P_{Y|Z}(y|z,\lambda_z)$ be families of distributions
and consider the adaptive GAMP algorithm, Algorithm \ref{algo:gamp},
run on a sequence of problems, indexed by the dimension $n$ satisfying
the following assumptions:
\begin{itemize}
\item[(a)] Same as Assumption \ref{as:gamp}(a) to (c) with $k=2$.  In addition,
the distributions for the vector $X$ is given by
$P_X(\cdot|\lambda_x^*)$ for some ``true"
parameter $\lambda_x^* \in \Lambda_x$ and the conditional distribution of $Y$ given
$Z$ is given by  $P_{Y|Z}(y|z,\lambda_z^*)$ for some ``true"
parameter $\lambda_z^* \in \Lambda_z$.

\item[(b)]  Same as Assumption \ref{as:agamp}(c).

\item[(c)]  The adaptation functions are set to \eqref{eq:HxML} and \eqref{eq:HzML}.
\end{itemize}
\end{assumption}

\begin{theorem} \label{thm:consistent}
Consider the outputs of the adaptive GAMP algorithm with ML adaptation as described in
Assumption \ref{as:agamp-ML}.
Then, for any fixed $t$,
\begin{itemize}
\item[(a)]  The components of $\theta_x^t$ and
$\theta_z^t$ in \eqref{eq:thetaxz}
converge empirically with bounded moments of order $k=2$ as in \eqref{eq:thetaLim}
and the limits \eqref{eq:taulamLim} hold almost surely.

\item[(b)] In addition, if $(\alpha_r^t,\xi_r^t) \in S_x(\tau_r^t)$,
and the family of distributions $P_X(\cdot|\lambda_x)$, $\lambda_x \in \Lambda_x$
is identifiable in Gaussian noise with parameter set $S_x(\tau_r^t)$
and pseudo-likelihood $\phi_x$ (see Definition~\ref{def:Pxident}), then
\beq \label{eq:lamxCons}
\lim_{n \arr \infty} \lambdahat^t_x = \lambdabar_x^t = \lambda^*_x
\eeq
almost surely.

\item[(c)]  Similarly, if $\Kbf_p^t \in S_z(\tau_p^t)$ for some $t$,
and the family of distributions $P_{Y|Z}(\cdot|\lambda_z)$,
$\lambda_z \in \Lambda_z$
is identifiable with Gaussian inputs with parameter set $S_z(\tau_p^t)$
and pseudo-likelihood $\phi_z$ (see Definition~\ref{def:Pzident}) then
\beq \label{eq:lamzCons}
\lim_{n \arr \infty} \lambdahat^t_z = \lambdabar_z^t = \lambda^*_z
\eeq
almost surely.
\end{itemize}
\end{theorem}
\emph{Proof:}  See Appendix \ref{sec:consistentPf}.

The theorem shows, remarkably, that for a very large class of the
parameterized distributions, the adaptive GAMP algorithm is able
to asymptotically estimate the correct parameters.  Moreover, there is asymptotically
no performance loss between the adaptive GAMP algorithm and a corresponding
oracle GAMP algorithm that knows the correct parameters in the sense that
the empirical distributions of the algorithm outputs are described by the same
SE equations.

There are two key requirements:
First, that the optimizations in \eqref{eq:HxML} and \eqref{eq:HzML} can be computed.
These optimizations may be non-convex.  Secondly, that the optimizations can be performed
are over
sufficiently large sets of Gaussian channel parameters $S_x$ and $S_z$ such that
it can be guaranteed that the SE equations eventually enter these sets.  In the examples
below, we will see ways to reduce the search space of Gaussian channel parameters.

\section{Numerical Results} \label{sec:numEx}

\subsection{Estimation of a Gauss-Bernoulli input} \label{sec:gaussBern}

Recent results suggest that there is considerable value in
learning of priors $P_X$ in the context of compressed sensing
\cite{CandesT:06}, which considers the estimation of sparse vectors $\xbf$
from underdetermined measurements ($m < n$) .
It is known that estimators such as LASSO offer certain optimal min-max performance over a large
class of sparse distributions \cite{DonJohMM:11}.
However, for many particular distributions, there is a potentially large
performance gap between LASSO and MMSE
estimator with the correct prior.  This gap was the main motivation for
\cite{VilaSch:11,VilaSch:12} which showed large gains of the EM-GAMP method
due to its ability to learn the prior.

Here, we illustrate the performance and asymptotic consistency
of adaptive GAMP in a simple compressed sensing example.
Specifically, we consider the
estimation of a sparse vector $\xbf \in \R^n$ from $m$ noisy measurements
\begin{equation*}
\ybf = \Abf \xbf + \wbf = \zbf + \wbf,
\end{equation*}
where the additive noise $\wbf$ is random with i.i.d.\ entries $w_i \sim \mathcal{N}(0, \sigma^2)$. Here, the ``output" channel is determined by the statistics on $\wbf$,
which are assumed to be known to the estimator.  So, there are no unknown
parameters $\lambda_z$.

As a model for the sparse input vector $\xbf$, we assumed the components are
i.i.d.\ with the Gauss-Bernoulli distribution,
\beq \label{eq:GaussBern}
    x_j \sim \left\{ \begin{array}{ll}
        0 & \mbox{prob } = 1-\rho, \\
        \mathcal{N}(0,\sigma_x^2) & \mbox{prob }  = \rho
        \end{array} \right.
\eeq
where $\rho$ represents the probability that the component is non-zero (i.e.\
the vector's sparsity ratio) and $\sigma_x^2$ is the variance of the
non-zero components.  The parameters $\lambda_x = (\rho, \sigma_x^2) $
are treated as unknown.

In the adaptive GAMP algorithm, we use the estimation functions
$G_x$, $G_s$, and $G_z$ corresponding to the sum-product GAMP
algorithm. As described in Appendix~\ref{sec:gampDetails}, for the sum-produce GAMP
the SE equations simplify so that $\alpha_r^t = 1$ and $\xi_r^t = \taubar_r^t$.
Since the noise variance is known, the initial output noise variance $\tau_r^0$ obtained
by adaptive GAMP in Algorithm~\ref{algo:gamp} exactly matches that of oracle GAMP. Therefore, for $t = 0$, the parameters $\alpha_r^t$ and $\xi_r^t$
do not need to be estimated, and~\eqref{eq:HxML} conveniently simplifies to
\begin{equation}
H_x(\rbf, \tau_r) = \argmax_{\lambda_x \in \Lambda_x} \left\{\frac{1}{n}\sum_{j = 1}^{n} \log p_R(r_j | \lambda_x, \tau_r)\right\},
\label{eq:simHxML}
\end{equation}
where $\Lambda_x = [0, 1] \times [0, +\infty)$. For iteration $t > 0$, we rely on asymptotic
consistency, and assume that the maximization~\eqref{eq:simHxML}  yields the correct parameter estimates, so that $\lambdahat_x^t = \lambda_x$. Then, in principle, for $t > 0$ adaptive GAMP uses the correct parameter estimates and we expect it to match the performance of oracle GAMP. In our implementation, we run EM update~\eqref{eq:lamEM} until convergence to approximate  the ML adaptation~\eqref{eq:simHxML}.

Fig.\ \ref{fig:mseresults} illustrates the performance of adaptive GAMP on
signals of length $n = 400$ generated with the parameters $\lambda_x = (\rho = 0.2, \sigma_x^2 = 5)$. The performance of adaptive GAMP is compared to that of LASSO with MSE optimal regularization parameter,
and oracle GAMP that knows the parameters of the prior exactly. For generating the graphs, we performed $1000$ random trials by
forming the measurement matrix $\Abf$ from i.i.d.\ zero-mean Gaussian random variables of variance
$1/m$. In Figure~\ref{fig:mseresults}(a), we keep the variance of the noise fixed to $\sigma^2 = 0.1$ and plot the average MSE of the reconstruction against the measurement ratio $m/n$. In Figure~\ref{fig:mseresults}(b), we keep the measurement ratio fixed to $m/n = 0.75$ and plot the average MSE of the
reconstruction against the noise variance $\sigma^2$. For completeness, we also provide the asymptotic MSE values computed via SE recursion.
The results illustrate that GAMP significantly outperforms LASSO over the whole range of $m/n$ and $\sigma^2$. Moreover, the results corroborate the consistency of adaptive GAMP which achieves nearly identical quality of reconstruction with oracle GAMP\@.
 The performance results indicate that adaptive GAMP can be an effective method for estimation when the parameters of the problem are difficult to characterize and must be estimated from data.

\subsection{Estimation of a Nonlinear Output Classification Function} \label{sec:nlClass}

As a second example, we consider the estimation of the linear-nonlinear-Poisson (LNP) cascade model~\cite{SchwartzPRS2006}. The model has been successfully used to characterize neural spike responses in early sensory pathways of the visual system. In the context of LNP cascade model, the vector $\xbf \in \R^n$ represents the linear filter, which models the linear receptive field of the neuron. AMP techniques combined with the parameter estimation have 
been recently proposed for neural receptive field estimation and 
connectivity detection in \cite{FletcherRVB:11}.

As in Section~\ref{sec:gaussBern}, we model $\xbf$ as a Gauss-Bernoulli vector of unknown parameters $\lambda_x = (\rho, \sigma_x^2)$. To obtain the measurements $\ybf$,  the vector $\zbf = \Abf \xbf$ is passed through a componentwise nonlinearity $u$ to result in
\beq
u(z) = \frac{1}{1 + \mathrm{e}^{-z}}.
\eeq
Let the function $\psi \in \R^r$ denote a vector
\beq
\psi(z) = \left(1, u(z), \dots, u(z)^{r-1}\right)^{T}.
\eeq
Then, the final measurement vector $\ybf$ is generated by a measurement channel with a conditional density of the form
\beq \label{eq:pyzExp}
    p_{Y|Z}(y_i|z_i,\lambda_z) = \frac{f(z_i)^{y_i}}{y!} \mathrm{e}^{-f(z_i)},
\eeq
where $f$ denotes the nonlinearity given by
\[
    f(z) = \exp\left(\lambda_z^{T}\psi(z)\right).
\]
Adaptive GAMP can now be used to also estimate vector of polynomial coefficients $\lambda_z$, which together with $\xbf$, completely characterizes the LNP system.

\begin{figure}[t]
\centering
	\includegraphics[width=8.5cm]{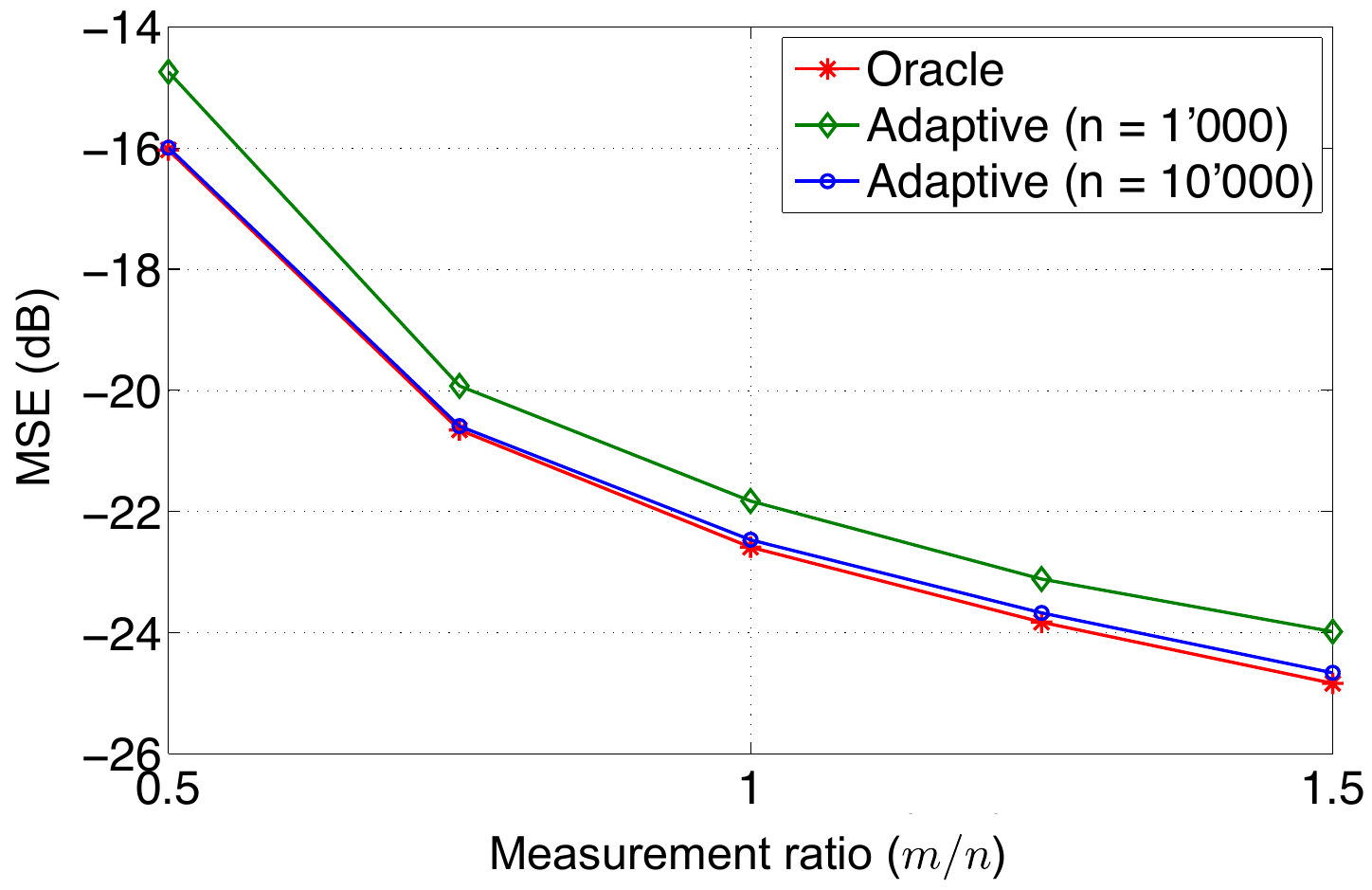}
	\caption{Identification of linear-nonlinear-Poisson cascade model. The average reconstruction MSE is plotted against the measurement ratio $m/n$. This plots illustrates near consistency of adaptive GAMP for large $n$.}
	\label{fig:nlresults}
\end{figure}

The estimation of $\lambda_z$ is performed with ML estimator described in Section~\ref{sec:MLadapt}. We assume that the mean and variance of the vector $\xbf$ are known at iteration $t=0$. This implies that for sum-product GAMP the covariance $\Kbf_p^0$ is initially known and the optimization~\eqref{eq:HzML} simplifies to
\begin{equation} \label{eq:simHzML}
H_z(\pbf, \ybf, \tau_p) = \argmax_{\lambda_z \in \Lambda_z} \left\{\frac{1}{m}\sum_{i = 1}^{m} \log p_Y(y_i | \lambda_z)\right\},
\end{equation}
where $\Lambda_z \subset \R^r$. The estimation of $\lambda_x$ is performed as in Section~\ref{sec:gaussBern}. As before, for iteration $t > 0$, we assume that the maximizations~\eqref{eq:simHxML} and~\eqref{eq:simHzML} yield correct parameter estimates $\lambdahat_x^t = \lambda_x$ and $\lambdahat_z^t = \lambda_z$, respectively. Thus we can conclude by induction that for $t > 0$ the adaptive GAMP algorithm should continue matching oracle GAMP for large enough $n$. In our simulations, we implemented~\eqref{eq:simHzML} with a gradient ascend algorithm and run it until convergence.

In Fig.~\ref{fig:nlresults}, we compare the reconstruction performance of adaptive GAMP against the oracle version that knows the true parameters $(\lambda_x, \lambda_z)$ exactly. We consider the vector $\xbf$ generated with true parameters $\lambda_x = (\rho = 0.1, \sigma_x^2 = 30)$. We consider the case $r = 3$ and set the parameters of the output channel to $\lambda_z = [-4.88, 7.41, 2.58]$. To illustrate the asymptotic consistency of the adaptive algorithm, we consider the signals of length $n = 1000$ and $n = 10000$. We perform $10$ and $100$ random trials for long and short signals, respectively, and plot the average MSE of the reconstruction against $m/n$. As expected, for large $n$, the performance of adaptive GAMP is nearly identical (within $0.15$) to that of oracle GAMP.

\section{Conclusions and Future Work}\label{sec:concl}

We have presented an adaptive GAMP method for the estimation of i.i.d.\ vectors $\xbf$ observed
through a known linear transforms followed by an arbitrary, componentwise random transform.
The procedure, which is a generalization of EM-GAMP methodology of \cite{VilaSch:11,VilaSch:12,KrzMSSZ:11-arxiv,KrzMSSZ:12-arxiv} that
estimates both the vector $\xbf$ as well as parameters in the source and
componentwise output transform.
In the case of large i.i.d.\ Gaussian transforms, it is shown that the
adaptive GAMP method is provably asymptotically consistent in that
the parameter estimates converge to the true values.  This convergence result holds over a large
class of models with essentially arbitrarily complex parameterizations.  Moreover,
the algorithm is computationally efficient since it reduces the vector-valued
estimation problem to a sequence of scalar estimation problems in Gaussian noise.
We believe that this method is applicable
to a large class of linear-nonlinear models with provable guarantees
can have applications in a wide range of problems.  We have mentioned the use of the method
for learning sparse priors in compressed sensing.  Future work will include learning
of parameters of output functions as well as possible extensions to non-Gaussian matrices.

\appendices

\section{Sum-Product GAMP Equations} \label{sec:gampDetails}

As described in \cite{Rangan:11-ISIT}, the sum-product estimation can be implemented with the estimation functions
\begin{subequations} \label{eq:Gsp}
\beqa
    G_x^t(r,\tau_r,\lambdahat_x) &:=&  \Exp[X|R=r,\tau_r,\lambdahat_x],
        \label{eq:Gxsp} \\
    G_z^t(p,y,\tau_p,\lambdahat_z) &:=&  \Exp[Z|P=p,Y=y,\tau_p,\lambdahat_z],
        \label{eq:Gzsp} \\
    G_s^t(p,y,\tau_p,\lambdahat_z) &:=&
        \frac{1}{\tau_p}\left( G_z^t(p,y,\tau_p,\lambdahat_z) - p \right),
\eeqa
\end{subequations}
where the expectations are with respect to the scalar random variables
\begin{subequations} \label{eq:GspDist}
\begin{align}
    R = X + V_x, \,\, &V_x \sim {\mathcal N}(0,\tau_r),
        \,\, X \sim P_X(\cdot|\lambdahat_x), \\
    Z = P + V_z, \,\, &V_z \sim {\mathcal N}(0,\tau_p),
        \,\, Y \sim P_{Y|Z}(\cdot|Z,\lambdahat_z).
\end{align}
\end{subequations}
\ifthenelse{\boolean{conference}}{}{
The paper \cite{Rangan:11-ISIT} shows that the derivatives of these estimation
functions for lines \ref{line:taus} and \ref{line:taux} are computed via the variances:
\begin{subequations} \label{eq:GspDeriv}
\begin{align}
    &\tau^r \frac{\partial G^t_x(r,\tau_r, \lambdahat_x)}{\partial r} = \var[X|R=r,\tau_r, \lambdahat_x] \\
    -&\frac{\partial G^t_s(p,y,\tau_p, \lambdahat_z)}{\partial p} \nonumber \\
    &=  \frac{1}{\tau_p}\left(
    1 - \frac{\var[Z|P=p,Y=y,\tau_p, \lambdahat_z]}{\tau_p}\right).
\end{align}
\end{subequations}
}%
The estimation functions \eqref{eq:Gsp}
correspond to scalar estimates of random
variables in additive white Gaussian noise (AWGN).
A key result of \cite{Rangan:11-ISIT} is that, when the
parameters are set to the true values (i.e.\ $(\lambdahat_x,\lambdahat_z) = (\lambda_x,\lambda_z)$),
the outputs $\xbfhat^t$ and $\zbfhat^t$ can be interpreted as sum products estimates
of the conditional expectations $E(\xbf|\ybf)$ and $E(\zbf|\ybf)$.
The algorithm thus reduces the vector-valued estimation problem
to a computationally simple
sequence of scalar AWGN estimation problems along with linear transforms.

Moreover, the SE equations in Algorithm~\ref{algo:SE} reduce to a particularly
simple forms, where $\taubar_r^t$ and $\xi_r^t$ in \eqref{eq:outSE} are given by
\begin{subequations}
\begin{equation}
\taubar_r^t = \xi_r^t =  \Exp^{-1}\left[\frac{\partial^2}{\partial p^2} \log p_{Y | P } (Y | P^t)\right],
\end{equation}
where the expectations are over the random variables $(Z,P^t) \sim {\mathcal N}(0,\Kbf_p^t)$
and $Y$ is given in \eqref{eq:YZPt}.  The covariance matrix $\Kbf_p^t$ has the form
\begin{equation}
\Kbf_p^t = \begin{bmatrix}
       \beta\tau_{x0} & \beta\tau_{x0} - \taubar_p^t      	\\[0.3em]
       \beta\tau_{x0} - \taubar_p^t & \beta\tau_{x0} - \taubar_p^t		\\[0.3em]
     \end{bmatrix},
\end{equation}
where $\tau_{x0}$ is the variance of $X$ and $\beta > 0$ is the asymptotic measurement ratio (see Assumption 1 for details). The scaling constant~\eqref{eq:alphaSE} becomes $\alpha_r^t = 1$.
The update rule for $\taubar_x^{\tp1}$ also simplifies to
\begin{equation}
\taubar_x^{\tp1} = \Exp \left[\var\left(X | R^t\right)\right],
\end{equation}
where the expectation is over the random variables in \eqref{eq:RVXt}.

\end{subequations}

\section{Convergence of Empirical Distributions} \label{sec:conv}

Bayati and Montanari's analysis in \cite{BayatiM:11}
employs certain deterministic
models on the vectors and then proves convergence properties
of related empirical distributions. To apply the same analysis here, we need
to review some of their definitions.  We say a function $\phi:\R^r \arr \R^s$
is \emph{pseudo-Lipschitz} of order $k>1$, if there exists an $L > 0$
such for any $\xbf$, $\ybf \in \R^r$,
\[
    \|\phi(\xbf) - \phi(\ybf)\|
        \leq L(1+\|\xbf\|^{k-1}+\|\ybf\|^{k-1})\|\xbf-\ybf\|.
\]

Now suppose that for each $n=1,2,\ldots$,
$\vbf^{(n)}$ is a set of vectors
\beq \label{eq:vsetApp}
    \vbf^{(n)} = \{\vbf_i(n), i=1,\ldots,\ell(n)\},
\eeq
where each element $\vbf_i(n) \in \R^s$ and $\ell(n)$ is the number
of elements in the set.  Thus, $\vbf^{(n)}$
can itself be regarded as a vector with $s\ell(n)$ components.
We say that $\vbf^{(n)}$
\emph{empirically converges with bounded moments of order $k$}
as $n\arr\infty$ to a random vector $\Vbf$ on $\R^s$ if:  For all
pseudo-Lipschitz continuous functions, $\phi$, of order $k$,
\[
    \lim_{n \arr\infty} \frac{1}{n} \sum_{i=1}^{n} \phi(\vbf_i(n))
    = \Exp(\phi(\Vbf)) < \infty.
\]
When the nature of convergence
is clear, we may write (with some abuse of notation)
\[
    \vbf^{(n)} \stackrel{\PL(k)}{\arr} \Vbf \mbox{ as } n \arr \infty,
\]
or
\[
      \lim_{n \arr \infty} \vbf^{(n)} \stackrel{\PL(k)}{=} \Vbf.
\]

Finally, let ${\mathcal P_k^s}$ be the set of probability distributions
on $\R^s$ with bounded $k$th moments, and suppose that
$H:{\mathcal P_k^s} \arr \Lambda$ is a functional
${\mathcal P_k^s}$ to some
topological space $\Lambda$.  Given a set $\vbf^{(n)}$ as in
\eqref{eq:vsetApp},
write $H(\vbf)$ for $H(P_{\vbf})$ where $P_{\vbf}$ is the
empirical distribution on the components of $\vbf$.  Also,
given a random vector $\Vbf$ with distribution $P_\Vbf$ write
$H(\Vbf)$ for $H(P_\Vbf)$.  Then, we will say that the functional
$H$ is \emph{weakly pseudo-Lipschitz continuous} of order $k$ if
\[
    \lim_{n \arr \infty} \vbf^{(n)} \stackrel{\PL(k)}{=} \Vbf
    \Longrightarrow
   \lim_{n \arr \infty} H(\vbf^{(n)}) = H(\Vbf),
\]
where the limit on the right hand side is in the topology
of $\Lambda$.

\section{Proof of Theorem \ref{thm:stateEvo}}
\label{sec:stateEvoPf}

The proof follows along the adaptation argument of \cite{RanganF:12-ISIT}.
We use the tilde superscript on quantities such
  as $\xorc^t, \rorc^t, \tauorc_r^t, \porc^t, \tau_p^t, \sorc^t$, and $\zorc^t$
to denote values generated via a non-adaptive version of the GAMP.
The non-adaptive GAMP algorithm has the
same initial conditions as the adaptive algorithm
(i.e.\ $\xorc^0 = \xbfhat^0, \tauorc_p^0 = \tau_p^0, \sorc^{-1} =\sbf^{-1} = 0$),
but with $\lambdahat_x^t$ and $\lambdahat_z^t$ replaced by their deterministic limits $\bar{\lambda}_x^t$ and $\bar{\lambda}_z^t$, respectively. That is, we replace
lines \ref{line:zhat}, \ref{line:shat} and \ref{line:xhat} with
\beqan
    \zorc_i^t &=& G_z^t(p_i^t,y_i,\tau_p^t,\lambdabar_z^t), \qquad
    \sorc_i^t = G_s^t(p_i^t,y_i,\tau_p^t,\lambdabar_z^t), \\
    \xorc_j^{\tp1} &=& G_x^t(r_j^t, \tau_r^t,\lambdabar_x^t).
\eeqan
This non-adaptive algorithm is precisely the standard GAMP method analyzed in
 \cite{Rangan:11-ISIT}.  The results in that paper show that the
 outputs of the non-adaptive algorithm satisfy all the required limits
 from the SE analysis.  That is,
\[
    \lim_{n \arr \infty} \tilde{\theta}_x^t \stackrel{\PL(k)}{=} \thetabar_x^t,
    \qquad
    \lim_{n \arr \infty} \tilde{\theta}_z^t \stackrel{\PL(k)}{=} \thetabar_z^t,
\]
where $\tilde{\theta}_x^t$ and $\tilde{\theta}_z^t$ are the sets
generated by the non-adaptive GAMP algorithm:
 \begin{align*}
    \tilde{\theta}_x^t := \left\{(x_j, \tilde{r}_j^t, \tilde{x}_j^{\tp1}): j = 1, \ldots, n\right\},
    \\
    \tilde{\theta}_z^t = \left\{(z_i, \tilde{z}_i^{t}, y_i, \tilde{p}_i^t): i = 1, \dots, m\right\}.
 \end{align*}

The limits \eqref{eq:thetaLim} are now proven through a continuity argument that shows that
the adaptive and non-adaptive quantities must asymptotically agree with one another.
Specifically,
we will start by proving that the following limits holds almost surely for all $t \geq 0$
\begin{subequations} \label{Eq:inputConvergence}
\beqa
    \lim_{n \rightarrow \infty} \Delta_x^t &=& \lim_{n \rightarrow \infty} \frac{1}{n}\|\xbfhat^t - \tilde{\xbf}^t\|^k_k = 0,\label{Eq:xconvergence},\\
    \lim_{n \rightarrow \infty} \Delta_{\tau_p}^t &=& \lim_{n \rightarrow \infty} |\tau^t_p - \tauorc^t_p| = 0\label{Eq:taupconvergence}
\eeqa
\end{subequations}
where $\|\cdot\|_k$ is usual the $k$-norm.
Moreover, in the course of proving~\eqref{Eq:inputConvergence},
we will also show that the following limits hold almost surely
\begin{subequations}
\label{Eq:convergences}
\begin{align}
&\lim_{m \rightarrow \infty} \Delta_p^t = \lim_{m \rightarrow \infty} \frac{1}{m} \|\pbf^t - \tilde{\pbf}^t \|_k^k = 0,\label{Eq:pconvergence}\\
&\lim_{n \rightarrow \infty} \Delta_r^t = \lim_{n \rightarrow \infty} \frac{1}{n} \|\rbf^t - \tilde{\rbf}^t \|_k^k = 0,\label{Eq:rconvergence}\\
&\lim_{m \rightarrow \infty} \Delta_s^t = \lim_{m \rightarrow \infty} \frac{1}{m} \|\sbf^t - \tilde{\sbf}^t \|_k^k = 0,\label{Eq:sconvergence}\\
&\lim_{m \rightarrow \infty} \Delta_z^t = \lim_{m \rightarrow \infty} \frac{1}{m} \|\zbfhat^t - \tilde{\zbf}^t \|_k^k = 0,\label{Eq:zconvergence}\\
&\lim_{n \rightarrow \infty} \Delta_{\tau_r}^t = \lim_{n \rightarrow \infty} |\tau^t_r - \tauorc^t_r| = 0,\label{Eq:taurconvergence}\\
&\lim_{n \rightarrow \infty} \lambdahat_x^t = \bar{\lambda}_x^t,\label{Eq:lambdaxconvergence}\\
&\lim_{n \rightarrow \infty} \lambdahat_z^t = \bar{\lambda}_z^t,\label{Eq:lambdazconvergence}
\end{align}
\end{subequations}
The proof of the limits~\eqref{Eq:inputConvergence} and~\eqref{Eq:convergences} is achieved by an induction on $t$.  Although we only need to show the above limits for $k=2$,
most of the arguments hold for arbitrary $k \geq 2$.  We thus present the general derivation
where possible.

To begin the induction argument, first note that the non-adaptive algorithm has the same initial conditions as the adaptive algorithm. Thus the limits~\eqref{Eq:inputConvergence} and~\eqref{Eq:sconvergence} hold for $t = 0$ and $t = -1$, respectively.

We now proceed by induction. Suppose that $t\geq0$ and the limits ~\eqref{Eq:inputConvergence} and~\eqref{Eq:sconvergence} hold for some $t$ and $t-1$, respectively. Since $\Abf$ has i.i.d.\ components with zero mean and variance $1/m$, it follows from the
Mar\v{c}enko-Pastur Theorem \cite{MarcenkoP:67} that
that its $2$-norm operator norm is bounded. That is, there exists a constant $C_A$ such that
\beq \label{eq:matrixBnd}
    \lim_{n \rightarrow \infty} \|\Abf\|_k \leq C_A,\,\, \lim_{n \rightarrow \infty} \|\Abf^{\mathrm{T}}\|_k \leq C_A.
\eeq
This bound is the only part of the proof that specifically requires $k=2$.
From \eqref{eq:matrixBnd}, we obtain
\begin{align}
    \lefteqn{ \|\pbf^t - \porc^t\|_k = \|\Abf\xbfhat^t - \tau_p^t\sbf^{\tm1} - \Abf\xorc^t + \tauorc_p^t\sorc^{\tm1}\|_k } \nonumber \\
    &= \|\Abf(\xbfhat^t  - \xorc^t) + \tau_p^t(\sorc^{\tm1} - \sbf^{\tm1}) + (\tauorc_p^t - \tau_p^t )\sorc^{\tm1}  \|_k \nonumber \\
&\leq \|\Abf(\xbfhat^t  - \xorc^t)\|_k + |\tau_p^t|\|\sorc^{\tm1} - \sbf^{\tm1}\|_k + |\tauorc_p^t - \tau_p^t| \|\sorc^{\tm1}  \|_k \nonumber \\
&\leqn{a} \|\Abf\|_k\|\xbfhat^t  - \xorc^t\|_k + |\tau_p^t|\|\sorc^{\tm1} - \sbf^{\tm1}\|_k + |\tauorc_p^{t} - \tau_p^{t}| \|\sorc^{\tm1}  \|_k\nonumber \\
&\leq C_A\|\xbfhat^t  - \xorc^t\|_k + |\tau_p^t|\|\sbf^{\tm1} - \sorc^{\tm1}\|_k + | \tau_p^t - \tauorc_p^t| \|\sorc^{\tm1}  \|_k \label{Eq:pbndA}
\end{align}
where (a) is due to the norm inequality $\|\Abf\xbf\|_k \leq \|\Abf\|_k\|\xbf\|_k$.
Since $k \geq 1$, we have that for any positive numbers $a$ and $b$
\beq \label{eq:pbndtwo}
    (a + b)^k \leq 2^k(a^k + b^k).
\eeq
Applying the inequality \eqref{eq:pbndtwo} into \eqref{Eq:pbndA}, we obtain
\begin{align}
    &\frac{1}{m}\|\pbf^t - \porc^t\|_k^k \notag\\
    &\leq \frac{1}{m}\left(C_A\|\xbfhat^t  - \xorc^t\|_k + |\tau_p^t|\|\sbf^{\tm1} - \sorc^{\tm1}\|_k + \Delta_{\tau_p}^t \|\sorc^{\tm1}  \|_k\right)^k
    \nonumber \\
    & \leq 2^kC_A\frac{n}{m}\Delta_x^t + 2^k|\tau_p^t|^k\Delta_s^{\tm1} + 2^k(\Delta_{\tau_p}^t)^k \left(\frac{1}{m} \|\sorc^{\tm1}  \|_k^k\right). \label{eq:pbndB}
\end{align}
Now, since $\sorc^t$ and $\tauorc_p^t$ are the outputs of the non-adaptive algorithm they satisfy the limits
\begin{subequations}
\begin{align} \label{eq:staupLim}
&\lim_{n \rightarrow \infty} \frac{1}{m}\|\sorc^t\|_k^k = \lim_{n \rightarrow \infty} \frac{1}{m}\sum_{i = 1}^m |\tilde{s}_i^t|^k = \Exp\left[|S^t|^k\right] < \infty, \\
&\lim_{n \rightarrow \infty} \tauorc_p^t = \taubar_p^t < \infty.
\end{align}
\end{subequations}
Now, the induction hypotheses state that $\Delta_x^t$, $\Delta_s^{\tm1}$ and
$\Delta_{\tau_p}^t \arr 0$.
Applying these induction hypotheses along the bounds \eqref{eq:staupLim},
and the fact that $n/m \rightarrow \beta$ we obtain~\eqref{Eq:pconvergence}.

To prove~\eqref{Eq:lambdazconvergence}, we first prove the empirical convergence
of $(\pbf^t,\ybf)$ to $(P^t,Y)$.  Towards this end, let $\phi(p,y)$ be any pseudo-Lipschitz
continuous function $\phi$ of order $k$.  Then
\begin{align}
\lefteqn{ \left|  \frac{1}{m} \sum_{i = 1}^m \phi(p_i^{t},y_i) -
    \Exp\left[\phi(P^t,Y)\right]\right| } \nonumber \\
    &\leq \frac{1}{m} \sum_{i = 1}^m \left|\phi(p_i^{t},y_i) -
        \phi(\tilde{p}_i^{t},y_i)  \right| \nonumber\\
        & + \left| \frac{1}{m} \sum_{i = 1}^m \phi(\tilde{p}_i^{t},y_i)
         -\Exp\left[\phi(P^t,Y)\right]\right|  \nonumber \\
    &\leqn{a} \frac{L}{m} \sum_{i = 1}^m \left(1 + |p_i^t|^{k-1} + |\tilde{p}_i^t|^{k-1}  + |y_i|^{k-1} \right)|p_i^t - \tilde{p}_i^t| \nonumber \\
    &+ \left| \frac{1}{m} \sum_{i = 1}^m \phi(\tilde{p}_i^{t},y_i) -\Exp\left[\phi(P^t,Y)\right]\right| \nonumber \\
    &\leqn{b} LC \Delta_p^t+\left| \frac{1}{m} \sum_{i = 1}^m \phi(\tilde{p}_i^{t},y_i) -\Exp\left[\phi(P^t,Y)\right]\right|. \label{eq:phipybnd}
\end{align}
In (a) we use the fact that $\phi$ is pseudo-Lipschitz, and
in (b) we use H\"older's inequality
$|\xbfhat^{\mathrm{T}}\ybf| = \|\xbf\|_k\|\ybf\|_q$ with $q = p/(p-1)$ and define $C$ as
\begin{align} \
    \lefteqn{ C:= \left[ \frac{1}{m}\sum_{i=1}^m \left(
    1 + |p_i^t|^{k-1} + |\tilde{p}_i^t|^{k-1}  + |y_i|^{k-1}\right) \right]^{k/(k-1)} } \nonumber \\
    &\leq \frac{1}{m}\sum_{i=1}^m \left(
    1 + |p_i^t|^{k-1} + |\tilde{p}_i^t|^{k-1}  + |y_i|^{k-1}\right)^{k/(k-1)} \nonumber \\
    &\leq \mbox{const}\x\left[ 1 +
    \left(\frac{1}{m}\left\|\pbf^t\right\|_k^k\right)^{\frac{k-1}{k}} \right. \nonumber\\
    &\left.+ \left(\frac{1}{m}\left\|\porc^t\right\|_k^k\right)^{\frac{k-1}{k}} +
    \left(\frac{1}{m}\left\|\ybf\right\|_k^k\right)^{\frac{k-1}{k}} \right],
    \label{eq:Cpbnddef}
\end{align}
where the first step is from Jensen's inequality.
Since $(\porc^t,\ybf)$ satisfy the limits for the non-adaptive algorithm we have:
\begin{subequations} \label{eq:ptildeybnd}
\begin{align}
    \lim_{n \rightarrow \infty} \frac{1}{m}\|\porc^t\|_k^k &=
     \lim_{n \rightarrow \infty} \frac{1}{m}\sum_{i = 1}^m |\tilde{p}_i^t|^k = \Exp\left[|P^t|^k\right] < \infty \\
    \lim_{n \rightarrow \infty} \frac{1}{m}\|\ybf\|_k^k &=
     \lim_{n \rightarrow \infty} \frac{1}{m}\sum_{i = 1}^m |y_i|^k
     = \Exp\left[|Y|^k\right] < \infty
\end{align}
\end{subequations}
Also, from the induction hypothesis~\eqref{Eq:pconvergence}, it follows that the adaptive output must satisfy the same limit
\begin{equation}
\label{Eq:PhatBound}
\lim_{n \rightarrow \infty} \frac{1}{m}\| \pbf^t\|_k^k = \lim_{n \rightarrow \infty} \frac{1}{m}\sum_{i = 1}^m |p_i^t|^k = \Exp\left[|P^t|^k\right] < \infty.
\end{equation}
Combining~\eqref{eq:phipybnd},
\eqref{eq:Cpbnddef}, \eqref{eq:ptildeybnd},~\eqref{Eq:PhatBound},~
\eqref{Eq:pconvergence} we conclude that for all $t \geq 0$
\beq \label{eq:pyconv}
    \lim_{n \rightarrow \infty} (\pbf^t,\ybf) \stackrel{\PL(k)}{=} (P^t,Y).
\eeq
The limit \eqref{eq:pyconv} along with \eqref{Eq:taupconvergence}
and the continuity condition on $H_z^t$ in Assumption \ref{as:gamp}(d)
prove the limit in~\eqref{Eq:lambdazconvergence}.

The limit~\eqref{Eq:pconvergence} together with continuity conditions on $G_z^t$ in Assumptions
 \ref{as:gamp}
 show that~\eqref{Eq:sconvergence},~\eqref{Eq:zconvergence} and~\eqref{Eq:taurconvergence} hold for $t$. For example, to show~\eqref{Eq:zconvergence}, we consider the limit $m \rightarrow \infty$ of the following expression
\begin{align*}
    \frac{1}{m}\|\zbfhat^{t} - \zorc^{t}\|_k^k &= \frac{1}{m}\|G_z^t(\pbf^t, \ybf, \tau_p^t, \lambdahat_z^t) - G_z^t(\porc^t, \ybf, \tau_p^t, \bar{\lambda}_z^t) \|_k^k\\
    &\leqn{a} \frac{L}{m} \|\pbf^t - \porc^t\|_k^k = L\Delta_p^t, 
\end{align*}
where at (a) we used the Lipschitz continuity assumption. Similar arguments can be used for~\eqref{Eq:sconvergence} and~\eqref{Eq:taurconvergence}.

To show~\eqref{Eq:rconvergence}, we proceed exactly as for~\eqref{Eq:pconvergence}.
Due to the continuity assumptions on $H_x$,
this limit in turn shows that~\eqref{Eq:lambdaxconvergence} holds almost surely. Then,~\eqref{Eq:xconvergence} and \eqref{Eq:taupconvergence} follow directly from the continuity of $G_x$ in Assumptions \ref{as:gamp},
together with~\eqref{Eq:rconvergence} and~\eqref{Eq:lambdaxconvergence}. We have thus shown that if the limits~\eqref{Eq:inputConvergence} and~\eqref{Eq:convergences} hold for some $t$, they hold for $t+1$. Thus, by induction they hold for all $t$.

Finally, to show~\eqref{eq:thetaLim}, let $\phi$ be any pseudo-Lipschitz continuous
function $\phi(x,r,\xhat)$, and define
\begin{align}
&\epsilon^t = \left|  \frac{1}{n} \sum_{j = 1}^m \phi(x_j, \tilde{r}_j^{t}, \tilde{x}_j^{\tp1}) -\Exp\left[\phi(X, R^t, \Xhat^{\tp1})\right]\right|,
\end{align}
which, due to convergence of non-adaptive GAMP, can be made arbitrarily small by choosing $n$ large enough. Then, consider
\begin{align}
&\left|  \frac{1}{n} \sum_{j = 1}^m \phi(x_j, \hat{r}_j^{t}, \xhat_j^{\tp1}) -\Exp\left[\phi(X, R^t, \Xhat^{\tp1})\right]\right| \nonumber\\
&\leq \epsilon_n^t + \frac{1}{n} \sum_{j = 1}^n \left|\phi(x_j, \hat{r}_j^{t}, \xhat_j^{\tp1}) - \phi(x_j, \tilde{r}_j^{t}, \tilde{x}_j^{\tp1})  \right| \nonumber\\
&\leqn{a} \epsilon_n^t + L\|\rbf^t - \rorc^t\|_1 + L\|\hat{\xbf}^{\tp1} - \xorc^{\tp1}\|_1
\nonumber\\
& + \frac{L^\prime}{n}\sum_{j = 1}^n \left(|\hat{r}_j^t|^{k-1} + |\tilde{r}_j^t|^{k-1} \right)(|\hat{r}_j^t - \tilde{r}_j^t| + |\xhat_j^{\tp1} - \tilde{x}_j^{\tp1}|)
\nonumber\\
& +  \frac{L^\prime}{n}\sum_{j = 1}^n \left(|\xhat_j^{\tp1}|^{k-1} + |\tilde{x}_j^{\tp1}|^{k-1}\right)(|\hat{r}_j^t - \tilde{r}_j^t| + |\xhat_j^{\tp1} - \tilde{x}_j^{\tp1}|) \nonumber\\
&\leqn{b} \epsilon_n^t + L\left(\Delta_r^t\right)^{\frac{1}{k}} + L\left(\Delta_x^t\right)^{\frac{1}{k}}\nonumber\\
&+ L^{\prime}\left(\Delta_r^t\right)^{\frac{1}{k}}\left((\tilde{M}_x^{\tp1})^{\frac{k-1}{k}}+(\hat{M}_x^{\tp1})^{\frac{k-1}{k}} + (\tilde{M}_r^{t})^{\frac{k-1}{k}}+(\hat{M}_r^{t})^{\frac{k-1}{k}}\right)\nonumber\\
&+ L^{\prime}\left(\Delta_x^t\right)^{\frac{1}{k}} \left((\tilde{M}_x^{\tp1})^{\frac{k-1}{k}}+(\hat{M}_x^{\tp1})^{\frac{k-1}{k}} + (\tilde{M}_r^{t})^{\frac{k-1}{k}}+(\hat{M}_r^{t})^{\frac{k-1}{k}}\right) \label{Eq:xbound}
\end{align}
where $L$, $L^{\prime}$ are constants independent of $n$ and
\begin{align*}
&\hat{M}_x^{\tp1} = \frac{1}{n}\left\|\hat{\xbf}^{\tp1}\right\|_k^k, \quad
\hat{M}_r^t = \frac{1}{n}\left\|\rbf^t\right\|_k^k, \\
&\tilde{M}_x^{\tp1} = \frac{1}{n}\left\|\tilde{\xbf}^{\tp1}\right\|_k^k, \quad
\tilde{M}_r^t = \frac{1}{n}\left\|\tilde{\rbf}^t\right\|_k^k
\end{align*}
In (a) we use the fact that $\phi$ is pseudo-Lipshitz, in (b) we use $\ell_p$-norm equivalence $\|\xbf\|_1 \leq n^{1-1/p}\|\xbf\|_k$ and H\"older's inequality $|\xbfhat^{\mathrm{T}}\ybf| = \|\xbf\|_k\|\ybf\|_q$ with $q = p/(p-1)$. By applying of~\eqref{Eq:xconvergence},~\eqref{Eq:rconvergence} and since, $\hat{M}_x^{\tp1}$, $\tilde{M}_x^{\tp1}$, $\hat{M}_r^{t}$, and $\tilde{M}_r^{t}$ converge to a finite value we can obtain the first equation of~\eqref{eq:thetaLim} by taking $n \rightarrow \infty$. The second equation in~\eqref{eq:thetaLim} can be shown in a similar way.
This proves the limits \eqref{eq:thetaLim}.

Also, the first two limits in \eqref{eq:taulamLim} are a consequence of
\eqref{Eq:lambdaxconvergence} and \eqref{Eq:lambdaxconvergence}.
The second two limits follow from continuity assumptions in Assumption \ref{as:gamp}(e)
and the convergence of the empirical distributions in \eqref{eq:thetaLim}.
This completes the proof.

\section{Proof of Theorem \ref{thm:consistent}}
\label{sec:consistentPf}

Part (a) of Theorem~\ref{thm:consistent} is a 
direct application of the general result, Theorem \ref{sec:stateEvo}.
To apply the general result, first observe that Assumptions~\ref{as:agamp-ML}(a)
and (c) immediately imply the corresponding 
items in Assumptions \ref{as:agamp}. So,
we only need to verify the continuity condition in Assumption \ref{as:agamp}(b)
for the adaptation functions in \eqref{eq:HxML} and \eqref{eq:HzML}. 

We begin by proving the continuity of $H_z^t$.
Fix $t$, and let $(\ybf^{(n)},\pbf^{(n)})$ be a
sequence of vectors and $\tau_p^{(n)}$ be a sequence of scalars such that
\beq \label{eq:yptauLim}
    \lim_{n \arr \infty} (\ybf^{(n)},\pbf^{(n)}) \stackrel{\PL(p)}{=} (Y,P^t)
    \qquad \lim_{n \arr \infty} \tau_p^{(n)} = \taubar^t_p,
\eeq
where $(Y,P^t)$ and $\taubar^t_p$
are the outputs of the  state evolution equations.
For each $n$, let
\beq \label{eq:lamhatCons}
    \lambdahat_z^{(n)} = H_z^t(\ybf^{(n)},\pbf^{(n)},\tau_p^{(n)}).
\eeq
We wish to show that $\lambdahat_z^{(n)} \arr \lambda_z^*$, the true
parameter.  Since $\lambdahat_z^{(n)} \in \Lambda_z$ and $\Lambda_z$ is compact,
it suffices to show that, any limit point of any convergent subsequence
is equal to $\lambda_z^*$.  So, suppose that $\lambdahat_z^{(n)} \arr \lambdahat_z$
to some limit point $\lambdahat_z$ on some subsequence $\lambdahat_z^{(n)}$.

From $\lambdahat_z^{(n)}$ and the definition \eqref{eq:HzML} 
it follows that
\begin{align} \label{eq:phizComp}
    \frac{1}{m} \sum_{i=1}^m & \phi_z(y_i^{(n)},p_i^{(n)},\tau_p^{(n)},\lambdahat_z^{(n)}) \nonumber\\
    &\geq \frac{1}{m} \sum_{i=1}^m \phi_z(y_i^{(n)},p_i^{(n)},\tau_p^{(n)},\lambda_z^*).
\end{align}
Now, since $\tau_p^{(n)} \arr \taubar^t_p$ and $\lambdahat_z^{(n)} \arr \lambdahat_z$,
we can apply the continuity condition in Definition~\ref{def:Pzident}(c) to obtain
\begin{align} \label{eq:phizCompB}
    \liminf_{n \arr \infty}
    \frac{1}{m} \sum_{i=1}^m &\left[ \phi_z(y_i^{(n)},p_i^{(n)},\taubar_p^t,\lambdahat_z)
    \right. \nonumber\\
    &\left.- \phi_z(y_i^{(n)},p_i^{(n)},\taubar_p^t,\lambda_z^*)
    \right] \geq 0.
\end{align}
Also, the limit \eqref{eq:yptauLim} and the fact that $\phi_z$ is
psuedo-Lipschitz continuous of order $k$ implies that
\beq \label{eq:phizCompC}
    \Exp[ \phi_z(Y,P^t,\taubar_p^t,\lambdahat_z) ]
    \geq \Exp[  \phi_z(Y,P^t,\taubar_p^t,\lambda_z^*) ].
\eeq
But, property (b) of Definition~\ref{def:Pzident}
shows that $\lambda_z^*$ is the maxima of the right-hand side, so
\beq \label{eq:phizCompD}
    \Exp[ \phi_z(Y,P^t,\taubar_p^t,\lambdahat_z) ] =
    \Exp[  \phi_z(Y,P^t,\taubar_p^t,\lambda_z^*) ].
\eeq
Since, by Definition~\ref{def:Pzident}(b), the maxima is unique, 
$\lambdahat_z = \lambda_z^*$.
Since this limit point is the same for all convergent subsequences, we see that
$\lambdahat^{(n)}_z \arr \lambda_z^*$ over the entire sequence.  We have thus shown that
given limits \eqref{eq:yptauLim}, the outputs of the adaptation function converge as
\[
    H_z^t(\ybf^{(n)},\pbf^{(n)},\tau_p^{(n)}) = \lambdahat_z^{(n)}  \arr
    \lambda_z^* = H_z^t(Y,P^t,\tau_p^{(n)}).
\]
Thus, the continuity condition on $H^t_z$ in Assumption \ref{as:agamp}(b) is satisfied.
The analogous continuity condition on $H^t_x$ can be proven in a similar manner.

Therefore, all the conditions of Assumption \ref{as:agamp} are satisfied and
we can apply Theorem \ref{thm:stateEvo}.  Part (a) of Theorem~\ref{thm:consistent} immediately follows from 
Theorem~\ref{thm:stateEvo}.

So, it remains to show parts (b) and (c) of Theorem~\ref{thm:consistent}.
We will only prove (b); the proof of (c) is similar.  Also, since 
we have already established \eqref{eq:taulamLim}, we only need to show that
the output of the SE equations  matches the true
parameter.  That is, we need to show $\lambdabar_x^t = \lambda_x^*$.
This fact follows immediately from the selection of 
the adaptation functions:
\beqa
    \lefteqn{ \lambdabar_x^t \stackrel{(a)}{=} H_x^t(R^t,\taubar^t_r) }
        \nonumber \\
    &\stackrel{(b)}{=}& 
    \argmax_{\lambda_x \in \Lambda_x}  \max_{(\alpha_r,\xi_r) 
        \in S_x(\taubar_r^t)}
        \Exp\left[  \phi_x(R^t,\lambda_x,\alpha_r,\xi_r) \right] 
            \nonumber \\
    &\stackrel{(c)}{=}&
    \argmax_{\lambda_x \in \Lambda_x}  
        \max_{(\alpha_r,\xi_r) \in S_x(\taubar_r^t)}
        \nonumber \\
    & &  \Exp\left[  \phi_x(\alpha_r^tX+V^t,\lambda_x,\alpha_r,\xi_r)
     | \lambda_x^*, \xi_r^t \right] \\
    &\stackrel{(d)}{=}& \lambdabar_x^* \label{eq:lamxLim}
\eeqa
where (a) follows from the SE equation \eqref{eq:lambarxSE};
(b) is the definition of the ML adaptation function 
$H_x^t(\cdot)$ when interpreted as a functional on a random variable $R^t$;
(c) is the definition of the random variable $R^t$ in \eqref{eq:RVXt}
where $V^t \sim {\mathcal N}(0,\xi_r^t)$; and
(d) follows from Definition~\ref{def:Pxident}(b) and 
 the hypothesis that $(\alpha_r^*,\xi_r^*) \in S_x(\taubar_r^t)$.
Thus, we have proven that $\lambdabar_x^t = \lambda_x^*$,
and this completes the proof of part (b) of Theorem \ref{thm:consistent}.
The proof of part (c) is similar.

\small{
\bibliographystyle{IEEEtran}
\bibliography{../bibl}
}

\end{document}